\documentclass[twocolumn]{aastex631}

\shorttitle{Energetic Particle Transport in the CME-disrupted Wind of AU~Microscopii}
\shortauthors{Fraschetti et al.}

\begin{document}

\title{Stellar Energetic Particle Transport in the Turbulent and CME-disrupted Stellar Wind of AU~Microscopii}

\correspondingauthor{Federico Fraschetti}
\email{federico.fraschetti@cfa.harvard.edu}

\author[0000-0002-5456-4771]{Federico Fraschetti}
\affil{Center for Astrophysics $|$ Harvard \& Smithsonian, 60 Garden Street, Cambridge, MA 02138, USA}
\affil{Dept. of Planetary Sciences-Lunar and Planetary Laboratory, University of Arizona, Tucson, AZ, 85721, USA}

\author[0000-0001-5052-3473]{Juli\'an D. Alvarado-G\'omez}
\affil{Leibniz Institute for Astrophysics Potsdam, An der Sternwarte 16, 14482 Potsdam, Germany}

\author[0000-0002-0210-2276]{Jeremy~J.~Drake}
\affil{Center for Astrophysics $|$ Harvard \& Smithsonian, 60 Garden Street, Cambridge, MA 02138, USA}

\author[0000-0003-3721-0215]{Ofer~Cohen}
\affil{University of Massachusetts at Lowell, Department of Physics \& Applied Physics, 600 Suffolk Street, Lowell, MA 01854, USA}


\author[0000-0002-8791-6286]{Cecilia Garraffo}
\affiliation{Center for Astrophysics \text{\textbar} Harvard \& Smithsonian, 60 Garden Street, Cambridge, MA 02138, USA}



\begin{abstract}
Energetic particles emitted by active stars are likely to propagate in astrospheric magnetized plasma turbulent and disrupted by the prior passage of energetic Coronal Mass Ejections (CMEs).  
We carried out test-particle simulations of $\sim$ GeV protons produced at a variety of distances from  the M1Ve star AU~Microscopii by coronal flares or travelling shocks. Particles are propagated within the large-scale quiescent three-dimensional magnetic field and stellar wind reconstructed from measured magnetograms, and {  within the same stellar environment following passage of a $10^{36}$~erg kinetic energy CME}. In both cases, magnetic fluctuations with an isotropic power spectrum are overlayed onto the large scale stellar magnetic field and particle propagation out to the two innnermost confirmed planets is examined. In the quiescent case, the magnetic field concentrates the particles onto two regions near the ecliptic plane. 
After the passage of the CME, the closed field lines remain inflated and the re-shuffled magnetic field remains highly compressed, shrinking the scattering mean free path of the particles. In the direction of propagation of the CME-lobes the subsequent EP flux is suppressed. Even for a CME front propagating out of the ecliptic plane, the EP flux along the planetary orbits highly fluctuates and peaks at $\sim 2 -3$ orders of magnitude higher than the average solar value at Earth, both in the quiescent and the post-CME cases. 
\end{abstract}



\section{Introduction} \label{sec:intro}

Active low mass K- and M-star circumstellar environments are affected with an occurrence rate much higher than Solar by violent eruptions producing either very energetic coronal flares \citep{Youngblood.etal:17,Jackman.etal:20} or possibly escaping Coronal Mass Ejections (CMEs, $ > 10^{31}$ erg kinetic energy) detected, e.g., via X-ray spectroscopy  \citep{Argiroffi.etal:19}; CME candidates are also traced via Doppler shift in Balmer lines \citep{Houdebine.etal:90} {  or asymmetries therein \citep{Vida.etal:19}}, continuous X-ray absorption during the flare \citep{Moschou.etal:19} {  or dimming in the extreme ultraviolet and X-ray due to the CME mass loss \citep{Veronig.etal:21}}. The broadband flare emission (from radio to $\gamma$-rays), hence the bolometric detectable energy output, from such stars is routinely investigated \citep[e.g.,][]{Paudel.etal:21} whereas the kinetic energy of the associated CMEs has been estimated only in a handful of cases \citep[e.g.,][]{Moschou.etal:19}. Within the heliosphere, the charged particle acceleration at CME-driven shocks has been accurately determined via in-situ measurements to drain $\sim 10\%$ of the total CME energy, regardless of the magnetic obliquity at the shock \citep{David.etal:22}. Comparable energy fractions might be expected at active stars. 

The passage of a CME compresses and breaks magnetic field lines leading to a re-arrangement of the large-scale magnetic field topology throughout the astrosphere, from the corona to the interplanetary region, that is traversed by charged particles energized close to the star. Such a disrupted configuration of the stellar wind is more likely to be encountered by outward propagating energetic particles (hereafter EPs) from active stars  
due to a flaring rate much higher than solar \citep{Youngblood.etal:17}. The flux of EPs onto habitable zone (hereafter HZ) planets in the quiescent winds of active stars was first determined numerically for the case of TRAPPIST-1 \citep{Fraschetti.Drake.etal:19}. The flux exceeded the solar value by $\sim 4$ orders of magnitude. However, the passage of a very energetic CME is expected to re-shuffle the wind magnetic field over large angular regions out to large distances. To our knowledge, the effect of such a phenomenon has not yet been investigated.

The James Webb Space Telescope (JWST) is expected to open up new pathways toward the observational studies of exoplanet habitability and atmosphere composition and  evolution. In particular, exoplanets with radii between $1.7$ and $3.5$ times the Earth radius (i.e., sub-Neptunes) are {  favorable targets for atmospheric observations instead of smaller planets as the larger amount of atmospheric $H_2$ acts as greenhouse gas allowing for stable liquid-water} \citep{Pierrehumbert.Gaidos:11,Hu.etal:21}. 

We focus here on AU Microscopii, an M dwarf with flaring activity observed by, e.g., the Hubble Space Telescope (HST) in far ultraviolet \citep{Redfield.etal:02} or XMM Newton in X-rays \citep{Magee.etal:03} and a modelled connection between flares \citep[Extreme Ultraviolet Explorer,][]{Cully.etal:94} and  ejected plasmoids self-similarly expanding in a CME fashion. The confirmation of two sub-Neptunian planets orbiting AU~Mic \citep{Martioli.etal:21} makes the system particularly attractive for investigating the effects of the CME passage on the EP propagation from star to planet due to their impact on planet atmosphere and its evaporation \citep{Fulton.etal:17}.

The diffusive transport of EPs originating from solar eruptions is known to be governed by the unperturbed large-scale magnetic field and by its small-scale fluctuations \citep{Jokipii:66}. 
Existing numerical analyses of the propagation of EPs from young stars surrounded by proto-planetary disks \citep{Rodgers-Lee.etal:17,Rab.etal:17,Fraschetti.Drake.etal:18,Padovani.etal:18,Gaches.Offner:18} or in exoplanetary environments \citep{Fraschetti.Drake.etal:19} {  have focused on quiescent stellar conditions.} 
Particle transport is determined by integrating EP trajectories {  in synthetic 3-dimensional turbulence \citep{Fraschetti.Drake.etal:19} or by solving a suitable transport equation, as done recently in \cite{Hu.etal:22}}. However, as mentioned above, EP propagation into a realistic astrosphere disrupted by a recent (within $1 - 2$ hours) CME passage does not appear to have been discussed previously.

In this paper, we perform a detailed analysis of the propagation of charged particles energized in proximity of AU Mic, i.e., by flares or CME-shocks, through the magnetized stellar wind {  calculated via the Space Weather Modeling Framework (SWMF) codes, in particular the Alfv\'en Wave Solar Model  \citep[AWSoM,][]{vanderHolst:14},} out to the second confirmed planet. {  Synthetic turbulent magnetic field is} added to the large scale unperturbed component \citep{Fraschetti.Drake.etal:19}. The propagation within the quiescent state astrosphere is compared with the propagation {  90 minutes} after the passage of a very energetic CME; a kinetic energy consistent with the best candidate event observed in this star so far ($\sim 10^{36}$~erg) is adopted \citep{Katsova.etal:99, Alvarado.etal:22}). 

The outline of this paper follows: in Sect. \ref{sec:windmod} the observational properties of the AM~Mic planetary system are summarized; in Sect.\ref{sec:environ} the assumptions on the stellar EP origin and propagation properties are emphasized along with the generated magnetic turbulence with intensity and injection scale as parameters. In Sect.\ref{sec:EPs_prop} the main results are presented for the cases of winds in quiescent state (with particles injected as close as the lower corona) and post-CME state. In Sect.\ref{sec:discussion} the EP fluxes impinging on the planets AU~Mic~b and ~c for the quiescent and post-CME case are compared; the {  EPs} transport properties for AU~Mic and TRAPPIST-1 are also compared; Sect.\ref{sec:conclusion} draws the conclusions of this work.

\section{The large-scale magnetized wind of AU~Mic }
\label{sec:windmod}

AU~Microscopii is a bright, nearby \citep[magnitude\footnote{http://simbad.u-strasbg.fr/simbad/sim-id?Ident=Au+Mic} $=8.6$, d=$9.72 \pm 0.04$ pc,][]{Gaia.DR2:18} M dwarf with mass $M_\star = 0.5 M_{\odot}$, radius $R_\star = 0.75 \,R_{\odot} =5.18 \times 10^{10}$~cm \citep{Plavchan.etal:20} and a rotation period of $4.85$~days  \citep{Torres.etal:72}. A spatially-resolved edge-on debris disk surrounds the star {  with a $\sim 50$ au inner radius} \citep{Kalas.etal:04}. Transiting Exoplanet Survey Satellite (TESS) light-curves confirmed that AU~Mic hosts a Neptune-size planet \citep[AU~Mic~b,][]{Plavchan.etal:20} with radius $ 1.05 \, R_N = 2.6 \times 10^9$~cm (where $R_N$  
is the Neptune radius), mass $ 1.00\, M_N$ (with $M_N$ 
is the Neptune mass), {  orbital distance} $R_b= 0.065$~au $= 19.1\, R_{\star} =9.89 \times 10^{11}$~cm \citep[orbits are assumed to be circular][]{Martioli.etal:21}, orbital period $8.46$ days \citep{Martioli.etal:21,Klein.etal:21b}, an inclination angle of magnetic field/stellar rotation axis $\sim 19^\circ$  \citep{Klein.etal:21b}, and an uncertain alignment between the spin axis of the host star and the orbital vector axis of the planet \citep{Addison.etal:21}. TESS revealed also a second orbiting Neptune-size planet (AU~Mic~c) with radius $ 0.84 \, R_N = 2.07 \times 10^9$~cm, mass $ 0.13\, M_N < M_c < 1.46\, M_N$, {  orbital distance} $R_c= 0.11$~au $= 29.\, R_{\star} =1.5 \times 10^{12}$~cm, orbital period $18.8$ days \citep{Martioli.etal:21}. The planetary orbital plane for both planets is located within 1 degree from the stellar equator 
\citep{Martioli.etal:21}. 

The 3D magnetized {  quiescent Stellar Wind (hereafter SW)} was computed using the Space Weather Modeling Framework codes \citep{Toth.etal:05,vanderHolst:14,Gombosi.etal:18}, evolved from the BATS-R-US MHD code \citep{Powell.etal:99} that was originally developed for the solar corona. The code {  uses as inner boundary condition} a magnetogram {  \citep[details in][]{Alvarado.etal:22}} describing the surface distribution of the radial magnetic field in the quiescent state and calculates the coronal heating and SW acceleration due to Alfv\'en wave turbulence dissipation, taking into account radiative cooling and electron heat conduction. The model has been validated with solar wind observations \citep{Cohen.etal:08b}; improvement of wave dissipation to electron and anisotropic proton heating lead to good agreement with magnetic field measured by Parker Solar Probe \citep{vanderHolst:22}. Further validations have been based on remote observations in solar minimum \citep{Sachdeva.etal:19} and maximum \citep{Sachdeva.etal:21} conditions. The code has been {  adapted and} used to simulate SWs and the space weather environments of exoplanets \citep[e.g.,][]{Vidotto.etal:15,Garraffo.etal:17,Cohen.etal:20,Evensberget.etal:21}, and has also been used to simulate CME eruptions from highly-magnetized stars \citep[e.g.,][]{Cohen.etal:11,Alvarado.etal:18,Alvarado.etal:19,Alvarado.etal:20,Alvarado.etal:22} {  or the associated radio emission {  in $\epsilon$ Eridani}  \citep{OFionnagain.etal:22}.} 

Stellar surface magnetic field distributions needed to drive stellar wind simulations have generally been based on Zeeman-Doppler Imaging (ZDI) observations. The AU~Mic large-scale B-field derived in such a way suffers some uncertainties. \cite{Kochukhov.Reiners:20} have shown that the use of distinct polarizations (circular and linear) from ESPaDOnS and HARPSpol instruments in the ZDI map leads to magnetic field strengths differing by a factor $10$ ($184$~G and $2$~kG, respectively). Using SPIRou polarization observations, \cite{Klein.etal:21b} found a $450$~G large-scale dipole field inclined by $\sim 20^{\circ}$ with respect to the rotation axis. Both maps were used in a companion paper \citep{Alvarado.etal:22} as an inner boundary to generate the 3D magnetized SW of AU~Mic, both in quiescent and CME-disrupted phase; {  in the present work, we have used the B-field produced in the cases 1 and 3 therein.} \cite{Cohen.etal:22}, using the same wind reconstruction, have determined the variations in Ly$\alpha$ absorption signatures during transits of AU Mic b due to the passage of a very energetic CME.
{  The \cite{Klein.etal:21b} maps are also implemented in \cite{Kavanagh.etal:21} to generate via AWSOM the AU Mic wind with $2$ distinct mass loss rates, that lead to different radio emission.}

\begin{figure}
	\includegraphics[width=8.5cm]{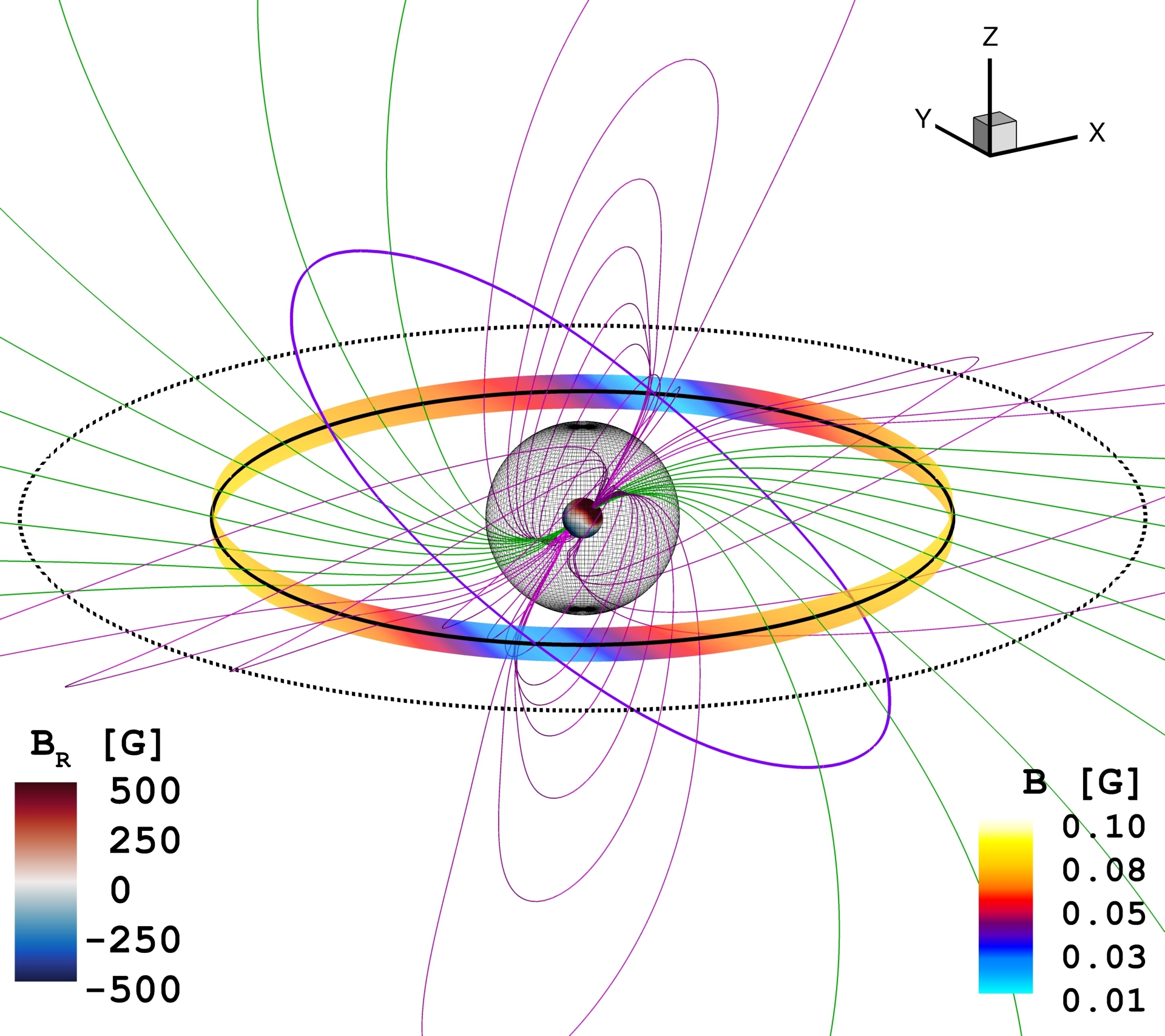}\\
\caption{A 3D view of the open (green) and closed (magenta) quiescent stellar wind magnetic field lines. The central sphere is the host star, color coded by the the radial component of the local stellar magnetic field on the surface. The transparent gridded sphere marks $R = 5\, R_{\star}$. The two circles on the equatorial plane mark the orbit of planet b (solid) and c (dotted) around the star. The colored circular ring at $R = R_b$ is color coded with the strength of the large scale magnetic field. The plane $B_r =0$ (where $B_r$ is the radial component of the magnetic field), that on average corresponds to the current sheet, is approximately denoted by the purple tilted circle.  
\label{fig:3D_UB}}
\end{figure}

In this paper, we calculate the propagation of stellar EPs within the turbulent magnetized SW of AU~Mic driven via the ZDI maps from \cite{Klein.etal:21b} and \cite{Kochukhov.Reiners:20}. In addition to the quiescent SW, we have produced a number of SW configurations disrupted by very energetic CMEs ({   kinetic energy $\sim 10^{36}$} erg, \cite{Alvarado.etal:22} propagating throughout the entire simulation box. 
{  The CME structure is initialized by using the \citet{Titov.Demoulin:99} flux-rope eruption model over the AWSoM field background.}
We consider herein one such configuration in detail: stellar EPs are propagated within SW snapshots $90$ minutes after the CME initialization, i.e., when the CME front has reached $>100\, R_\star$ region{  . EPs travel much faster than the CME front and  the choice of $90$ minutes ensures that the astrosphere has been disrupted by the CME as far as the numerical calculation allows.}

\section{Stellar energetic particles in the turbulent environment of AU~Mic} \label{sec:environ}
\subsection{Assumptions for EPs: origin, propagation and abundance}\label{sec:assumptions}

The goal of this work is to compare the transport of EPs in a quiescent SW environment of a young active star with the transport within the same environment $90$ minutes after the passage of an extraordinarily energetic CME (compared with heliospheric scales). 

In the solar context, GOES measurements of proton enhancements at $1$~AU confirm that EPs might originate both from SXR flares and CME-driven shocks \citep{Belov.etal:07}. Guided by the heliospheric observations, we assume that young active stars produce EPs via two distinct processes: 1) CME-driven shocks, travelling through the interplanetary medium and therein accelerating EPs, a certain fraction of which are likely to escape the shock at various distances from the host star; 2) coronal flares that release EPs locally accelerated close to the stellar surface. Presumably via different mechanisms, namely diffusive shock acceleration for the former and magnetic reconnection for the latter, both processes contribute  multi-MeV to $\sim$~GeV kinetic energy protons in the heliosphere. 

We inject EPs on spherical surfaces with radius $R_s = 2\, R_\star$ and $5\, R_\star$ concentric with the star, {  to compare the effect of the injection at two locations where the relative amount of closed-to-open field lines changes with a large spatial gradient; the diffusion in that region determines the EP flux at the planets.} Following the approach in \cite{Fraschetti.Drake.etal:19}, we calculate the time-forward propagation of test particles by using two distinct magnetostatic SW configurations: 1) quiescent interplanetary magnetic field; 2) stellar magnetic field 90 minutes after the initiation of a very energetic CME; in both cases we include the same overlapping small scale turbulence (see Sect. \ref{sec:postCME_SW}). 

The modelling of TESS flaring rate for AU Mic \citep{Gilbert.etal:21} is consistent with 1 flare every 3.8 hours for a flux between 0.06$\%$ and 1.5$\%$ of the stellar flux  \citep{Martioli.etal:21}. For the AU~Mic bolometric luminosity of 0.09 $L_{\odot}$ \citep{Plavchan.etal:09}, 
these correspond to fluxes at AU~Mic~b of $1.8 \times 10^4$ {  erg cm$^{-2}$ s$^{-1}$ sr$^{-1}$} and $4.4 \times 10^5$ {  erg cm$^{-2}$ s$^{-1}$ sr$^{-1}$} and, possibly, to very energetic associated CMEs.  
Thus, the 90-minutes {  interval} elapsed since the CME initialization allows the CME front to reach $R > 100 \, R_\star$ before the eruption of a subsequent energetic CME, {  consistently with observations \citep{Gilbert.etal:21}.}  
On the other hand, due to the high flaring occurrence in active stars, i.e., a likely high rate of associated CMEs, a wind configuration disrupted by a CME has a higher filling factor than for the Sun; therefore a significant fraction of EPs originating from the host star encounters typically a non-quiescent wind. 
In contrast, due to the lower level of solar activity, in the heliosphere most of the EP transport occurs within a quiescent, rather than a CME-disrupted, wind.

The magnetostatic approximation adopted herein is justified as follows. The MHD wind solution and the magnetic turbulence are stationary on the time-scale of EP propagation to a good approximation. The EPs travel close to the speed of light whereas the stellar rotation period of $4.86$ d and radius $0.75 \, R_{\odot}$ \citep{Klein.etal:21b} imply a surface stellar rotation speed of a few km~s$^{-1}$; the Alfv\'en  speed in the circumstellar is at most a few thousand km~s$^{-1}$ throughout the simulation box,  
which is much smaller than the EPs speed.  

The EP abundance in the circumstellar medium at a given distance from the host star cannot be constrained through direct observation; instead, we use the estimate based on solar scaling relations between EP fluence and far-UV and SXR fluence during flares by \citet{Youngblood.etal:17}{  ; see Sect.\ref{sec:discussion} below.} This scaling provides a time-averaged EP enrichment for time scales comparable with a statistically typical flare duration \citep{Vida.etal:17}. {  We have verified that a total number of EPs $N_{inj} = 10,240$ yields numerical convergence in all cases presented herein.}

{  Finally, we note that the AU Mic detected debris disk is not expected to impact the EP transport as the inner radius of the disk is measured to be 50 au \citep{Kalas.etal:04}, which is much greater than $R_c$.
}

\subsection{Turbulent stellar magnetic field}\label{sec:turbulence}

Leveraging the universality of the Kolmogorov scaling within the turbulent inertial range \citep{Armstrong.etal:95}, we assume that the magnetic fluctuations around AU~Mic support a 3D Kolmogorov isotropic power spectrum \citep[see also][]{Fraschetti.Drake.etal:19}.
Spectral analy\-sis of Parker Solar Probe (PSP) measurements near the minimum of the solar cycle 25 in the quiescent inner heliosphere (as close as $0.2$ AU to the Sun) have shown \citep{Zhao.Zank.etal:20} that the power spectrum of the magnetic turbulence in the direction aligned with the magnetic field is consistent with a Kolmogorov spectral slope of $-5/3$ and in tension with the $-2$ slope predicted by the critical balance conjecture   \citep{Goldreich.Sridhar:95}; in addition, the perpendicular transport in the \cite{Goldreich.Sridhar:95} was found to be  inefficient in the perpendicular diffusion of fast particles \citep{Fraschetti:16a,Fraschetti:16b}. The spectral slope found by PSP confirms previous findings at 1 AU from the {\it Wind} spacecraft, restricted to fast solar wind \citep{Telloni.etal:19}, and a number earlier analyses \cite[e.g., ][]{Jokipii.Coleman:68}. 

The total magnetic field is decomposed as  
${\bf B(x) = B}_0 ({\bf x}) + \delta {\bf B(x)}$,
where the large-scale component ${\bf B}_0 ({\bf x})$ is the 3D magnetic field generated by the 3D-MHD wind simulations (see Section~\ref{sec:windmod}); the random component ${\bf \delta B} = {\bf \delta B} (x, y, z)$ has a zero mean ($\langle \delta {\bf B(x)} \rangle = 0$). As for the turbulent environment of TRAPPIST-1 \citep{Fraschetti.Drake.etal:19}, the fluctuation ${\bf \delta B} (x, y, z)$ is calculated as the sum of plane waves with random orientation, polarization, and phase following the prescription in \citet{Giacalone.Jokipii:99,Fraschetti.Giacalone:12}, with an inertial range $k_{\rm min} < k < k_{\rm max}$, with $k_{\rm max}/k_{\rm min} = 10^2$, where $k_{\rm max}$ is the magnitude of the wavenumber corresponding to a turbulence dissipation scale. 

The advantage of the test-particle approach used here is that particle trajectories enable tracking of the pitch-angle scattering, of the perpendicular diffusion, and also of the transport across field-lines both in the quiescent and CME-disrupted winds; such effects are known to contribute significantly to particle transport in the heliosphere \citep[e.g.,][]{Droege.etal:10,Fraschetti.Jokipii:11,Gomez-Herrero.etal:15} but are often neglected for analytic tractability. Moreover, a 1D-spatial transport equation approach cannot be applied to a wind disrupted by the CME passage as the radial  scaling of the diffusion coefficient, $\kappa$, typically inferred from the radial scaling of the large scale magnetic field, is reshuffled in the post-CME wind: the expected  strong angular dependence of $\kappa$ cannot be included in a semi-analytic model. 

A second parameter of the stellar wind magnetic turbulence is the correlation length $L_c$, i.e., the outer scale of turbulence injection. Due to the lack of observational constraints on $L_c$, and the likely observational inaccessibility to $L_c$ in the near future, 
in \cite{Fraschetti.Drake.etal:19} we used for TRAPPIST-1 a range of values of $L_c$, each one kept uniform throughout the simulation box, and found no significant difference in the spatial distribution of EPs at the distances corresponding to the semi-major axes of the planets in that system. Likewise, for AU~Mic we adopt here the uniform value $L_c =10^{-5}$~AU throughout the simulation box. Such a value warrants that the resonant scattering condition holds with good approximation during the EP propagation throughout the wind. The resonance condition reads ${\rm k} r_g({\mathbf x})/2\pi = r_g({\mathbf x})/L_c < 1$ for each wave-number ${\rm k}$ within the inertial range; here, $r_g ({\mathbf x}) = p_\perp c/ e B_0 ({\mathbf x})$ is the gyroradius of a proton with $p_\perp$ momentum perpendicular to the unperturbed and space-dependent magnetic field $B_0 ({\mathbf x})$, $e$ is the proton electric charge and $c$ is the speed of light in vacuum.
As for the case of TRAPPIST-1, the combined effect of a high surface stellar magnetic field strength and its decrease with radius make $L_c =10^{-5}$ AU a reasonable value within the assumed circular orbits of AU Mic b and c, for the particle energies considered. 

The power of the magnetic fluctuation $\delta B ({\bf x})$ relative to $B_0 ({\bf x})$ is defined as 
\begin{equation}
\sigma^2 = (\delta B ({\mathbf x}) /B_0 ({\mathbf x}))^2 .
\label{sigma}
\end{equation}
Given the current lack of any observational constraint of the magnetic turbulence around AU~Mic, it seems reasonable to assume a uniform $\sigma^2$, following \citet{Fraschetti.Drake.etal:18}. 
Here, $\sigma^2$ is assumed to be independent of space throughout the simulation box. The solar wind measurements between $0.3$ and $4$ AU yield for the turbulence amplitude $\delta B$ a power-law dependence on heliocentric distance with a comparable slope ($-2.2$) at a variety of helio-latitudes \citep{Horbury.Tsurutani:01}. In the steady-state reconstructed 3D magnetic fields used here (see Sect.\ref{sec:windmod}), the spherical average of the unperturbed field $\langle B_0 ({\bf x}) \rangle_{\Omega}$ drops with radius $R$ as $\sim R^{-2.2}$ \citep[see also the case of TRAPPIST-1 in][]{Fraschetti.Drake.etal:19}. The high anisotropy of the post-CME stellar wind due to the CME eruption causes deviations from the monotonic scaling of $\langle B_0 ({\bf x}) \rangle_{\Omega}$, but it is conceivable that the level of small-scale turbulence is not significantly altered (see Sect.\ref{sec:postCME_SW} and \cite{Kilpua.etal:21}). 

The turbulence within the young and active M dwarf magnetosphere is likely to be much stronger than in the solar wind \citep[$\sigma^2 \lesssim 0.1$, ][]{Burlaga.Turner:76}, hence the broader  $\sigma^2$-range $0.01 - 1.0$ is spanned here. The interpretation of our simulations makes use of the  scattering mean free path, $\lambda_\parallel$, given by quasi-linear theory \citep{Jokipii:66}, that reads \citep{Giacalone.Jokipii:99,Fraschetti.Drake.etal:18}
\begin{equation}
\lambda_\parallel ({\mathbf x}) \simeq 4.8 (r_g({\mathbf x})/L_c)^{1/3} L_c/\sigma^2 \, .
\label{eq:lambda}
\end{equation}
The choices of uniform $L_c$ and $\sigma^2$ imply that $\lambda_\parallel$ depends on spatial coordinates only via $r_g ({x})$, i.e., $B_0 ({x})$. \cite{FitzAxen.etal:21} investigated the transport of $< 1$ GeV protons within protostellar cores by implementing scattering off magnetic turbulence via a Monte Carlo algorithm that neglects perpendicular transport; {  thus, cross-field diffusion and consequent longitudinal spread cannot be incorporated. It is noteworthy} that assigning $L_c$ and $\sigma^2$ for a given particle energy (namely $\lambda_\parallel$) does not uniquely describe the particle transport due to the increase of the perpendicular diffusion as $\sigma^2$ increases \citep{Giacalone.Jokipii:99,Fraschetti.Giacalone:12,Droege.etal:16}. Therefore the EP trajectory needs to be calculated step by step in the given magnetic turbulence {  without assuming that EPs follow the magnetic field lines}.

\section{Results}
\label{sec:EPs_prop}

Here, we discuss the results from injecting $0.1$ and $1$ GeV kinetic energy protons
at $R_s = 2 R_{\star} = 0.0069$~au and at $R_s = 5 R_{\star} = 0.017$~au for $\sigma^2 = 0.01$, $1.0$. EPs are propagated in our simulations throughout the astrosphere until either they collapse back to the star or hit (for the first time along their trajectory) the spherical surface at $R = R_b$ and $R = R_c$. A small fraction of EPs hit at first the $R_b$-sphere at a latitude different from the geometrical cross-section of the planetary orbit, then backscatter and in their subsequent star-ward propagation hit the $R_b$-($R_c$-) sphere again at {  the latitude of the planet orbital plane}. Such a fraction is $< 1 \%$ at most, so is neglected here and {  EP trajectories are followed within the region $R < R_c$.}

\subsection{Quiescent stellar wind}
\label{sec:quiesc_SW_dist}

\begin{figure*}
	\includegraphics[width=9.2cm]{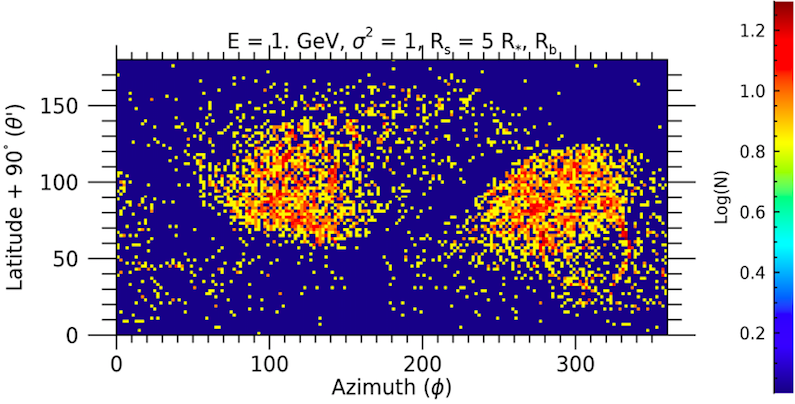}
	\includegraphics[width=9.2cm]{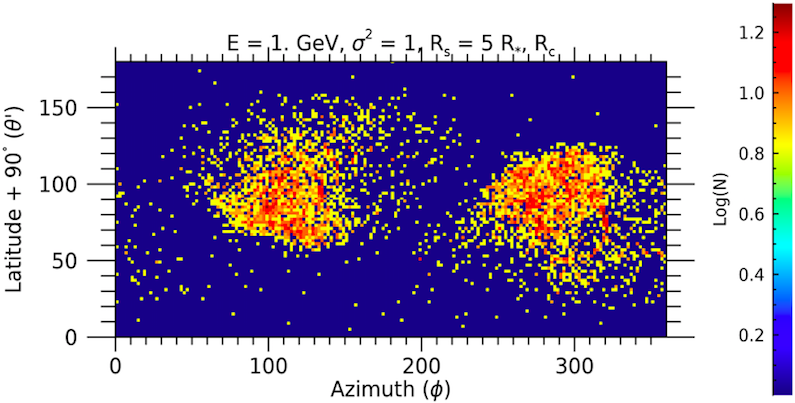}
	\caption{2D histograms in spherical coordinates (in degrees) of the hitting points of 1~GeV kinetic energy protons injected at $R_s = 5 R_{\star}$ into the quiescent stellar wind solution constructed from the ZDI radial field map from \cite{Klein.etal:21b} for $\sigma^2 = 1$. The polar angle is defined as $\theta'= \theta + 90^\circ$, where $\theta$ is the latitude centered in the star. The panels correspond to spherical surfaces at  
	{  $R_b$ (left) and $R_c$ (right).} Here, $L_c = 10^{-5}$ au. {  The planets orbital plane  corresponds here to a horizontal line $\theta'=  90^\circ$.}
	The log-scale colorbar indicates the number of EPs within each $2^{\circ} \times 2^{\circ}$ pixel. The same total number of EPs was injected in each case shown below. \label{fig:E1_varB1_Rs5}
	}
\end{figure*}

\begin{figure*}
	\includegraphics[width=8.7cm]{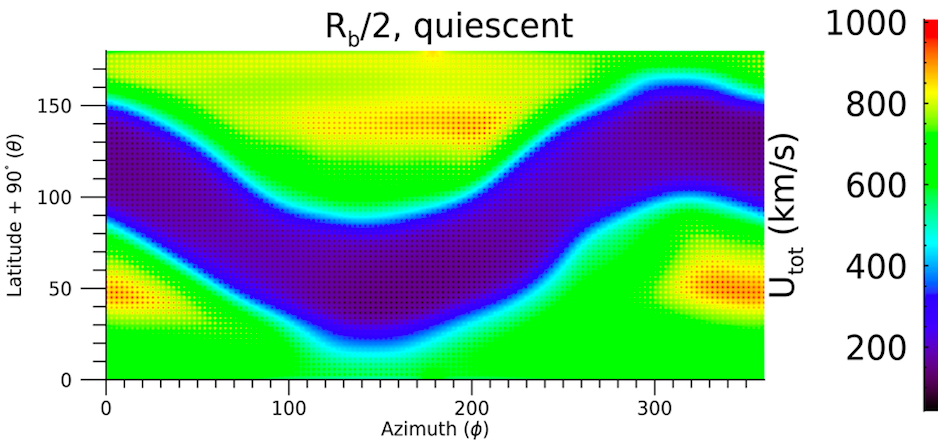}
	\includegraphics[width=8.7cm]{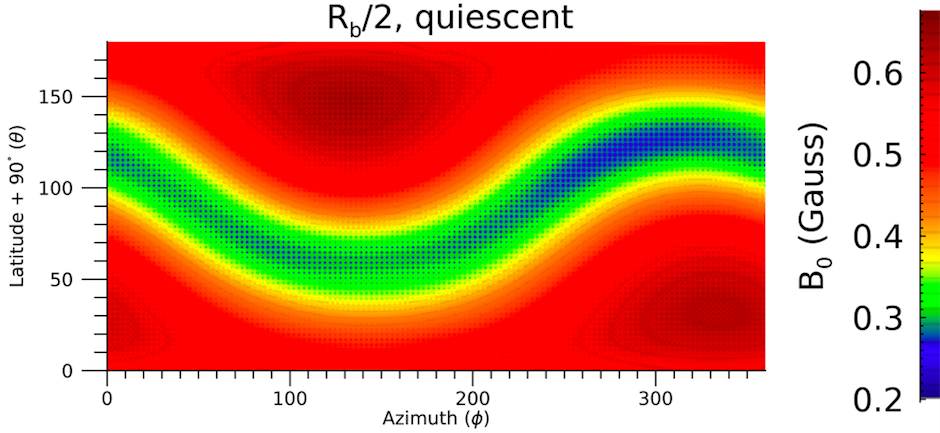}\\
	\includegraphics[width=8.7cm]{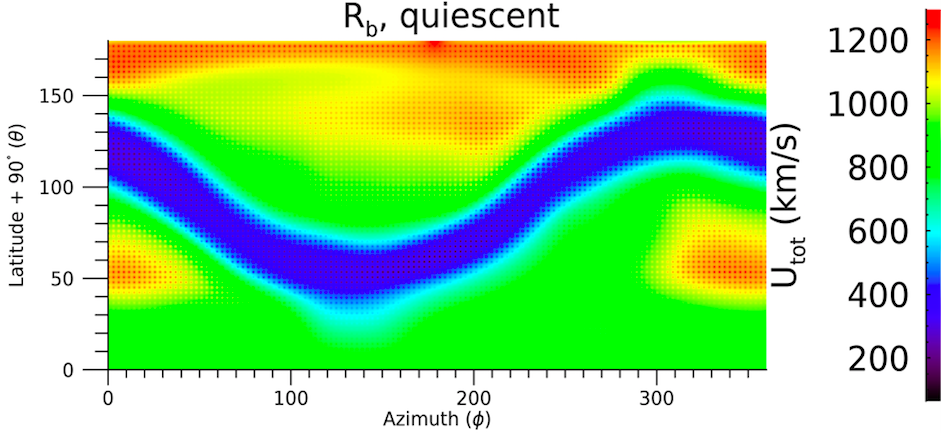}
	\includegraphics[width=8.7cm]{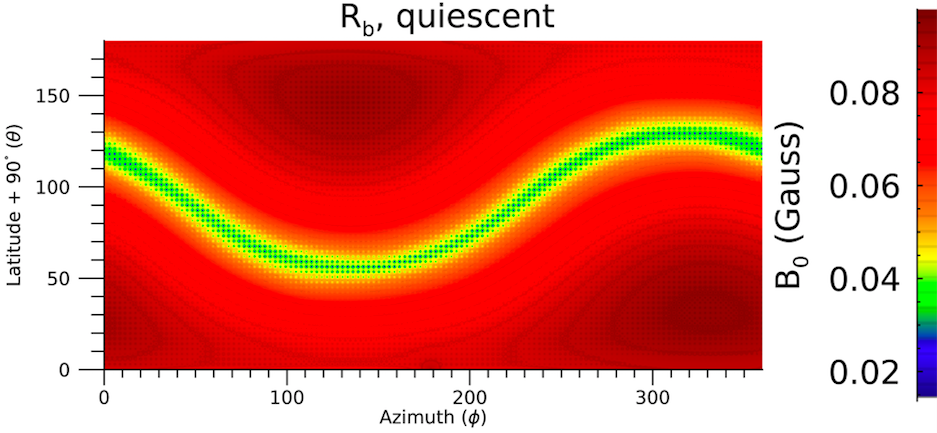}\\
    \includegraphics[width=8.7cm]{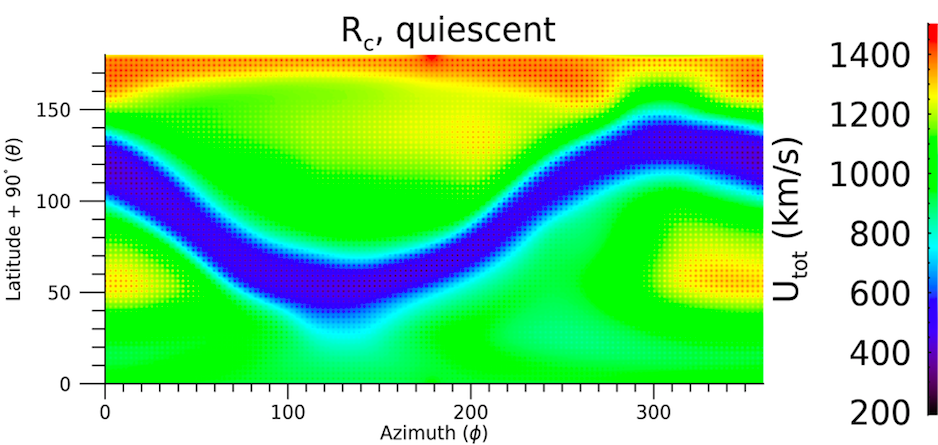}
	\includegraphics[width=8.7cm]{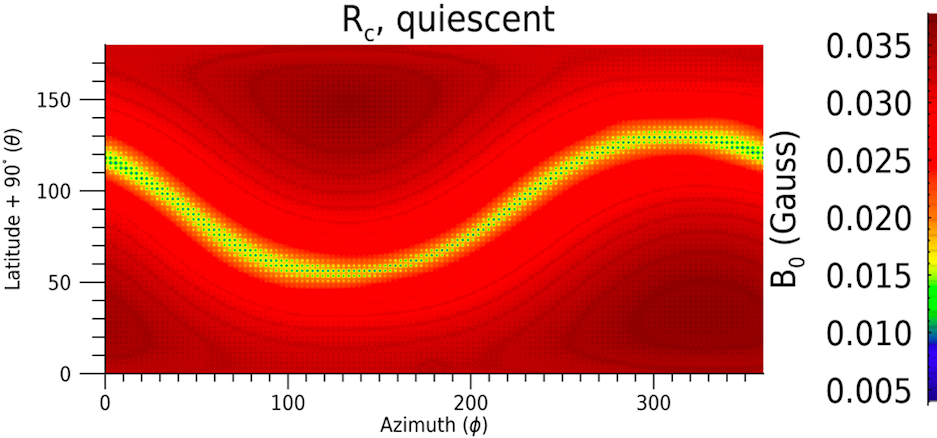}\\
\caption{Top row: The total wind flow speed, $U$, (left) and magnitude of the unperturbed magnetic field, $B_0$, (right) in the quiescent wind solution driven by the ZDI map from \cite{Klein.etal:21b} on the spherical surface at $R=R_b/2$. 
Middle row: same as top row at $R=R_b$. Bottom row: same as top row at $R=R_c$. 
\label{fig:2D_BU}}
\end{figure*}

\begin{figure*}
    \includegraphics[width=8.8cm]{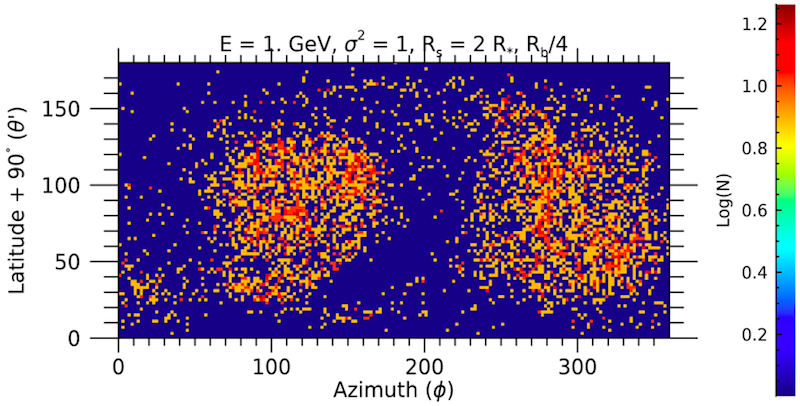}
	\includegraphics[width=8.8cm]{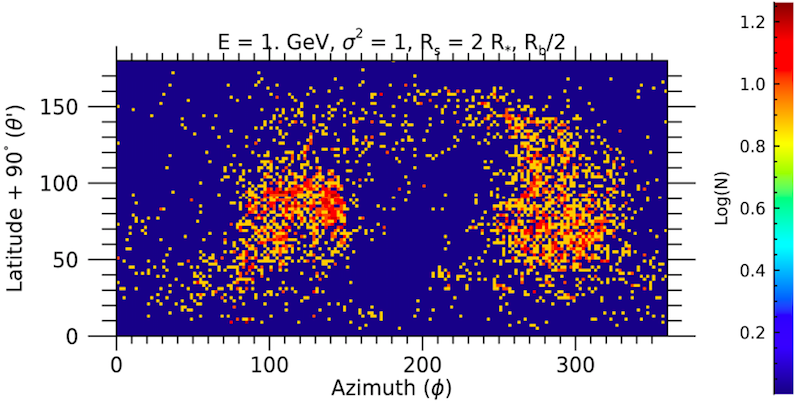}\\
	\includegraphics[width=8.8cm]{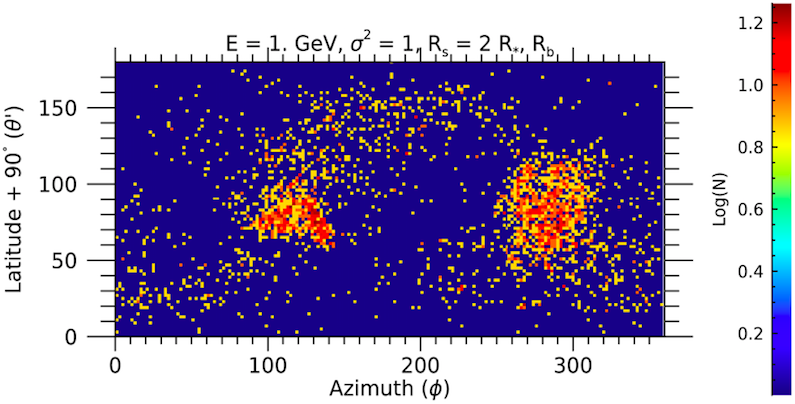}
	\includegraphics[width=8.8cm]{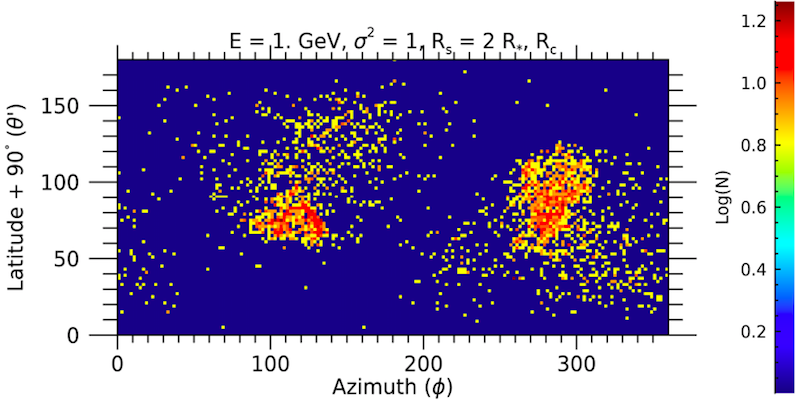}
\caption{Same as Fig. \ref{fig:E1_varB1_Rs5}, except with particles injected closer in, at $R_s = 2 R_{\star}  = 1.04 \times 10^{11}$ cm $= 0.0069$ au. The panels corresponds to spherical surfaces at distinct radii: 
$ 0.25 \,R_b = 4.8 \, R_{\star} = 0.0165$ au (top left), $0.5 \,R_b = 0.033 {\rm AU}$ (top right), $R_b = 0.066$~au (bottom left) and $R_c = 0.11$~au (bottom right). 
\label{fig:E1_varB1_Rs2}}
\end{figure*}

\begin{figure*}
	\includegraphics[width=8.8cm]{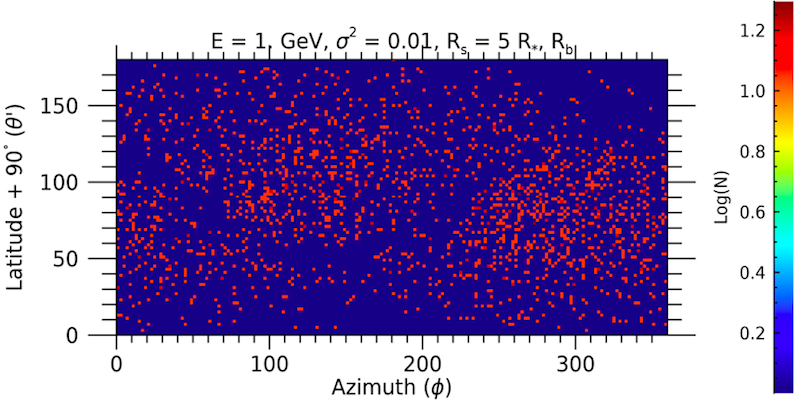}
	\includegraphics[width=8.8cm]{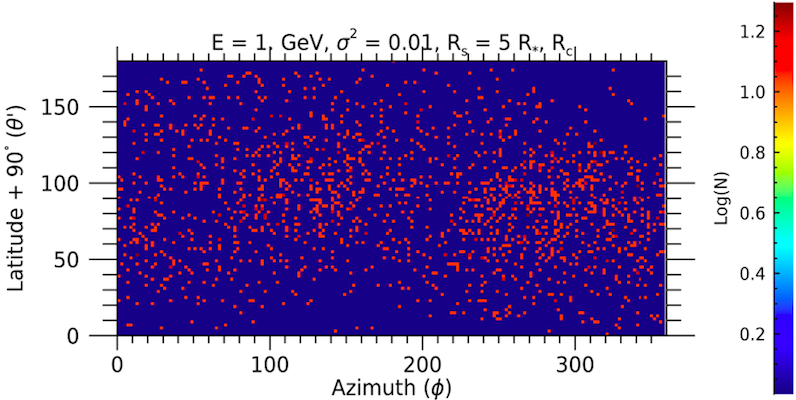}
\caption{The same as Fig. \ref{fig:E1_varB1_Rs5} except for a weaker turbulence of $\sigma^2 = 0.01$. 
\label{fig:2D_E_1d3_rs5_varB0p01}}
\end{figure*}

In Fig.~\ref{fig:3D_UB} open/closed magnetic field lines are seeded through the orbits of planets b and c. 
The unperturbed magnetic field lines that approximately track the motion of EPs have a {  predominantly dipolar structure \citep{Alvarado.etal:22}}.

For the quiescent stellar wind numerically reconstructed from the ZDI map from \cite{Klein.etal:21b}, Figure~\ref{fig:E1_varB1_Rs5} shows the 2D histogram in spherical coordinates at two distinct radii ($R = R_b$ and $R_c$) of the first-crossing points for 1~GeV kinetic energy protons in the case of strong turbulence, i.e., $\sigma^2 = 1$, injected at $R_s = 5 R_{\star}$. The corresponding stellar wind magnetic field is shown in Fig.\ref{fig:3D_UB}.

Depleted regions (in blue) reached by no EPs are found at both distances $R_b$ and $R_c$. Such regions broaden progressively as the distance from the host star increases. The azimuthal oscillation of the depleted regions maps the slow speed wind and the stellar current sheet, both shown in 2D spherical projections of the flow speed and the magnetic field strength in Fig.~\ref{fig:2D_BU}; similar correspondence between the EP 2D histogram and the current sheet was found in the case $\sigma^2 = 1$ for the HZ of the TRAPPIST-1 system  \citep{Fraschetti.Drake.etal:19}. As for TRAPPIST-1, the depleted regions result from the perpendicular transport, enhanced in the case of $\sigma^2 = 1$, of EPs at the boundary between open and closed field lines that favours a net migration of EPs across the large scale magnetic field from open to closed lines, due to the larger scattering mean free path in the weaker B-field of that region (see Eq.\ref{eq:lambda}): after transferring onto a closed line, EPs precipitate in a nearly scatter-free regime along the closed lines toward the star surface. Due to the larger B-field (smaller mean free path at fixed $\sigma^2$) in the open lines region, the number of EPs migrating via perpendicular transport in the opposite direction, from closed to open, is smaller, as some can backscatter and return to a closed line. Along the open lines the EPs proceed in an outward trajectory toward the planet. The effect of a weaker turbulence ($\sigma^2 = 0.01$) is discussed below in this section. 

As found in the case of TRAPPIST-1 \citep{Fraschetti.Drake.etal:19}, for AU-Mic the EP-depleted regions are explained by the combined effect enhanced perpendicular diffusion and B-dependence of $\lambda_\parallel$. The planet AU-Mic b is further out from the host star than TRAPPIST-1e, the closest HZ planet therein, i.e., 0.066 au compared with 0.029 au, but closer in units of stellar radius, $19\, R_\star$ compared with $52\, R_\star$; however, such differences do not lead to significant differences in the 2D histogram. 

Figure \ref{fig:E1_varB1_Rs2} shows that, if EPs are injected further in ($R_s=2\,R_{\star}$), the larger fraction of closed-to-open magnetic field lines in the inner wind increases the likelihood for EPs to be captured by the closed lines, hence a smaller EP flux at the planet. {  The upper panels show the decrease with radius of EPs due to the trapping by closed lines.}  

The combination of open field lines and strong turbulence focusses EPs into caps far out, {  allowing to reach the planetary ecliptic}. These caps track the high B-field projected region (see Fig.\ref{fig:2D_BU}).   
The high B-field shrinks the particle gyroscale, keeping them confined within a limited angle defining the caps. The magnetic field-rotation axis inclination angle of $19^\circ$ \citep{Klein.etal:21b} causes the caps to be tilted toward the equatorial plane. Likewise, a magnetic field/rotation axis inclination angle of $\sim 40^\circ$ in TRAPPIST-1 focusses EP caps toward the equatorial plane \citep{Fraschetti.Drake.etal:19}]. In case of alignment of the B-field and rotation axes, the projection of the current sheet would appear closer to an horizontal stripe in Fig.\ref{fig:2D_BU} and the EP caps would be expected to be closer to the polar region. 

The caps cross the  
equatorial plane, i.e., planetary orbit \citep{Plavchan.etal:20}, implying a modulation in the bombardment of the planet and in the consequent atmospheric ionization rate. Likewise, a modulation of the EP flux at the HZ planet TRAPPIST-1e was found to be up to $\sim 4 - 5$ orders of magnitude greater than experienced by Earth  \citep{Fraschetti.Drake.etal:19}, and its implications for the EP penetration depth were outlined in \citep{Fraschetti.Drake.proc.etal:21}. 

Also of interest is the timescale of particle modulation due to orbital motion and stellar rotation. The rotation period of AU~Mic is shorter than the orbital period of the planets \citep[8.46 days for AU Mic b][]{Klein.etal:21b}. The stellar rotation relative to the orbital motion will then sweep the EP caps over the planet with an effective period of approximately 11 days. The change in EP flux from EP-depleted to EP-enhanced regions occurs over an azimuthal angle range of a few 10s of degrees, such that the EP flux variation timescale would be of the order of a day. This is relevant for the recovery timescale of a planetary atmosphere to EP ionization events, and whether or not the atmosphere would be in a perturbed equilibrium state or subject to strong secular variation \citep{Herbst.etal:19,Chen.etal:21}.

The azimuthal variation of the EP flux impinging on the planet along its orbit around the star also strongly depends on the strength of the magnetic turbulence, as can be seen by comparing the case of strong (Fig.\ref{fig:E1_varB1_Rs5}) and weak turbulence (Fig.\ref{fig:2D_E_1d3_rs5_varB0p01}). In the former case, the planet crosses two discernible caps, whereas in the latter the planet orbits into an azimuthally homogeneous{ , but much more sparse,} distribution of EPs with far lower EP flux, evident also from the uniform color distribution in the latter.  
In the weak turbulence case, the distribution is closer to homogeneity as a result of the homogeneous injection on the $R_s$-sphere and the boundary between closed and open field lines does not act as an attractor for EPs toward the stellar surface as in the case $\sigma^2 = 1$; a comparably homogeneous distribution was found in the case of the TRAPPIST-1 system  \citep{Fraschetti.Drake.etal:19}. The difference between the distributions in Figs.~\ref{fig:2D_E_1d3_rs5_varB0p01} and \ref{fig:E1_varB1_Rs5} emphasizes the role of the diffusion in the direction perpendicular to the large-scale unperturbed field that is not accounted for in a purely Monte-Carlo approach to scattering off the magnetic fluctuations.

\begin{figure*}
\includegraphics[width=8.8cm]{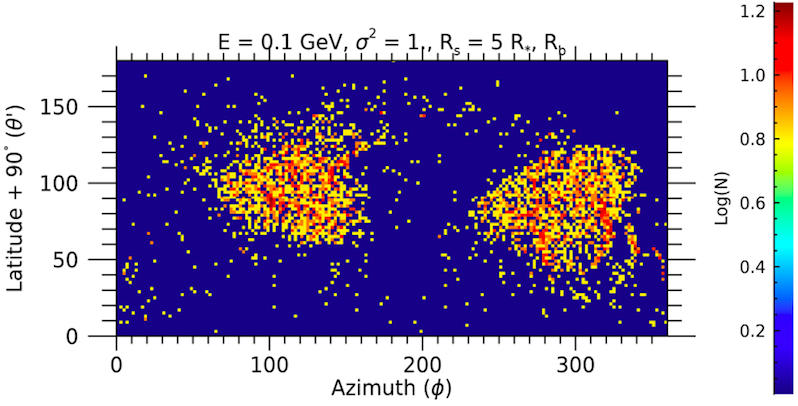}
\includegraphics[width=8.8cm]{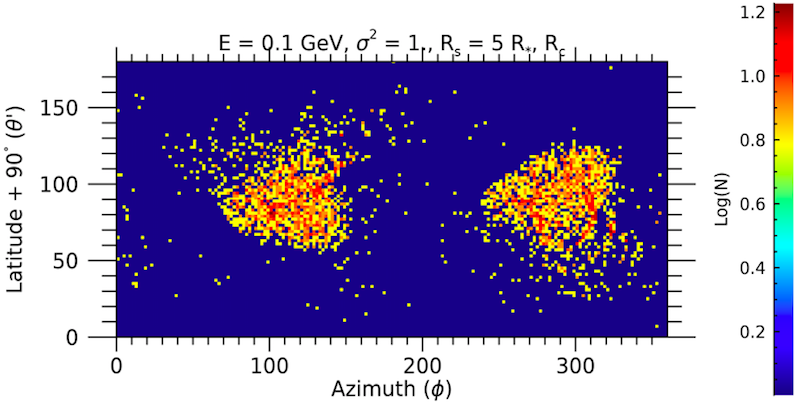}
\caption{The same as Fig.\ref{fig:E1_varB1_Rs5} except for $E = 0.1 $ GeV.
\label{fig:E_0p1_varB1}}
\end{figure*}

\subsection{Energy dependence of particle propagation}
\label{sec:EP_spectrum}

The spatial distribution of EPs is fairly independent of particle energy, as shown by comparing the 2D histograms for $0.1$ GeV EPs (see Fig. \ref{fig:E_0p1_varB1}) with $1$ GeV (see Fig.\ref{fig:E1_varB1_Rs5}) at two distinct radii, for the same injection radius $R_s = 5 R_\star$. {  As a consequence an energy-dependence of the diffusion coefficient does not alter the EPs 2D histograms.} A comparison of the EP energy spectrum at various astrospheric distances requires spanning a wide range of EP energies, in order to include the effect of the perpendicular transport that solar in-situ measurements suggest contribute significantly to circumsolar events \citep{Fraschetti.Jokipii:11,Gomez-Herrero.etal:15}. This work in hand focuses on the spatial distribution of EPs throughout the astrosphere to investigate the effect of the magnetic connection source-planet on the EP propagation. Transport might steepen the momentum spectrum of EPs at high energy, as shown with a pure scattering model with no perpendicular diffusion \citep{Li.Lee:15}; however, spectral modifications are not investigated herein because the source of EPs is not localized to an individual shock with specified parameters, i.e., fixed spectral shape.

\subsection{EP injection by flares in the stellar corona}
\label{sec:flare_corona}

\begin{figure*}
\includegraphics[width=8.8cm]{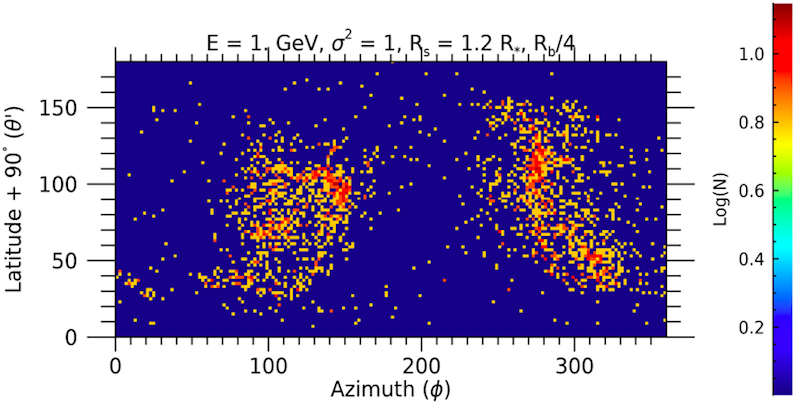}
\includegraphics[width=8.8cm]{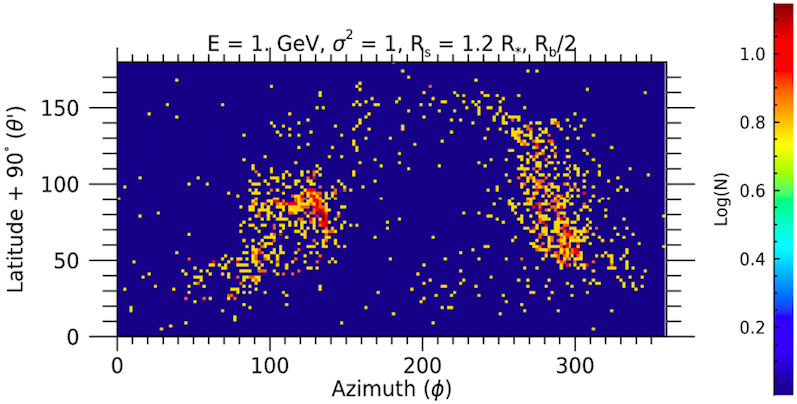}\\
\includegraphics[width=8.8cm]{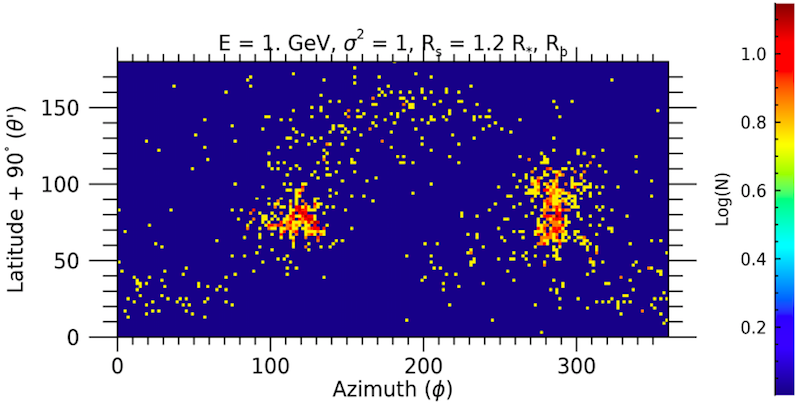}
\includegraphics[width=8.8cm]{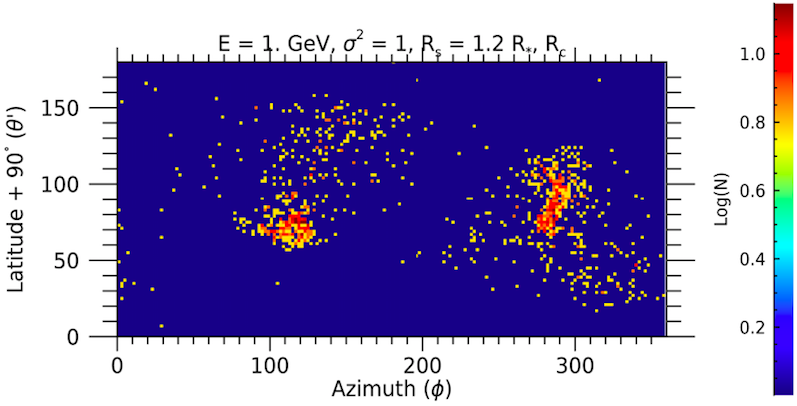}
\caption{Same as Fig.\ref{fig:E1_varB1_Rs5} for $R_s = 1.2 R_{\star}$ projected at $R = 0.25 \, R_b$ (left top panel) $R = 0.5 \, R_b$ (right top panel), at $R = R_b$ (left bottom panel) and at $R = R_c$ (right bottom panel). 
\label{fig:E_1d3_varB_1d0_Flare}}
\end{figure*}

Although the structure of a flaring loop within a high-resistivity stellar corona cannot be produced by our ideal MHD simulations, we can mimick the flare-produced EPs by releasing them within 1 $R_\star$ from the stellar surface. Figure \ref{fig:E_1d3_varB_1d0_Flare} depicts for the quiescent SW the 2D histogram at four distinct locations ($R = 0.25 \, R_b$, $R = 0.5 \,R_b$, $R_b$ and $R_c$) of EPs injected within the lower corona ($R_s = 1.2 R_{\star}$), 
that represents flare-emitted EPs. The same EPs pattern as at larger injection radius $R_s$ is found, including depleted regions, as wide as $\Delta \phi \sim 100^\circ$. In the coronal region the magnetic field suppresses $\lambda_\parallel$ by about a factor of $10$, favouring EPs precipitation to the star from the open/closed lines boundary, as discussed above. The EP flux to the planet is as intense as for greater $R_s$, although is limited over two smaller azimuthal intervals, i.e., $\Delta \phi \sim 40^\circ$ or $\sim 1$ (or $2$) days along planet b (c) orbits. 

\subsection{Post-CME stellar wind}
\label{sec:postCME_SW}

\begin{figure*}
    \includegraphics[width=9.1cm]{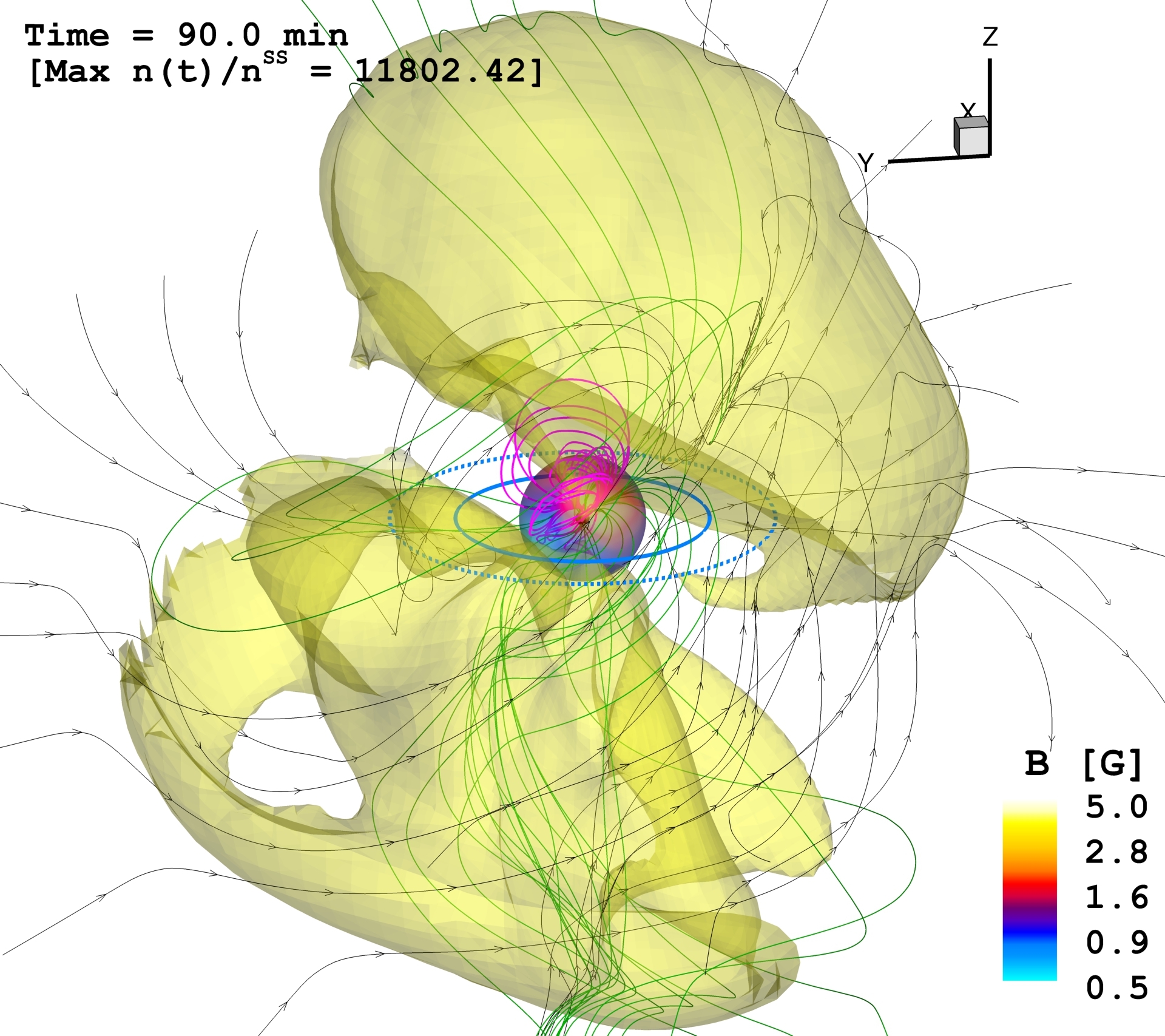}
    \includegraphics[width=9.1cm]{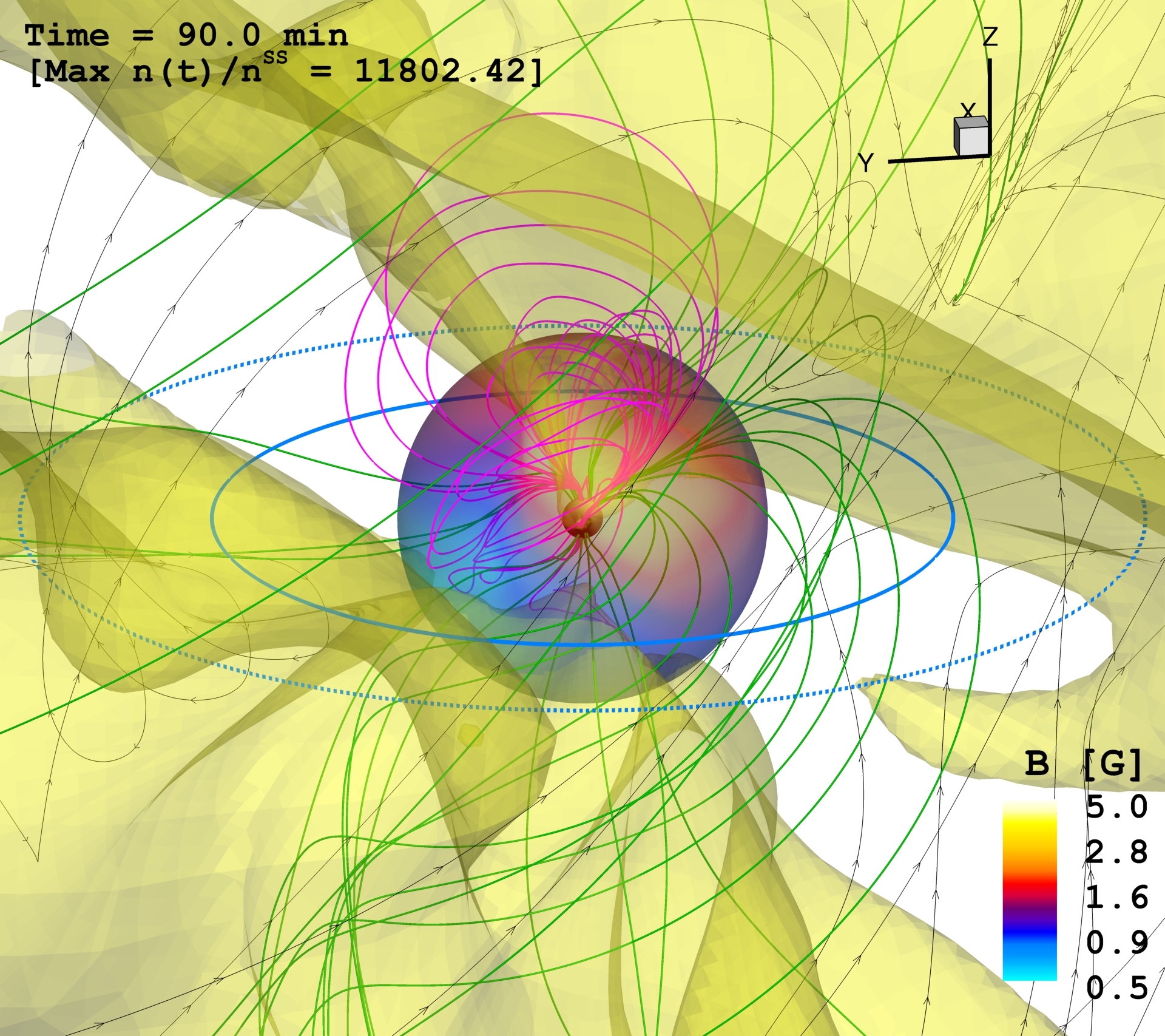}
    \caption{{\bf Left} Snapshot in a $175\, R_\star$ field of view of the bi-lobed plasma density isosurface ($n(\mathbf{x},t)/n_{SS}(\mathbf{x}) = 10.0$, where $n(\mathbf{x},t)$ is the density at location $\mathbf{x}$ and time $t$ and $n_{SS}(\mathbf{x})$ is the steady-state density at that location) of the propagating CME within the wind reconstructed from ZDI maps in \cite{Klein.etal:21b} at a time 90 minutes past the CME initiation. The sphere at $R_b/2$ is color coded by the B-field strength. The two cyan circles represent the orbits of planet b (solid) and c (dotted). The magenta lines indicate selected closed magnetic field lines. Open magnetic field lines are in green  if one end is attached to the star surface and in  black if none of the ends is tracked, respectively. {\bf Right} Same as the Left panel with a $60\, R_\star$ field of view. \label{fig:3D_CME} }
\end{figure*}

\begin{figure*}
	\includegraphics[width=8.8cm]{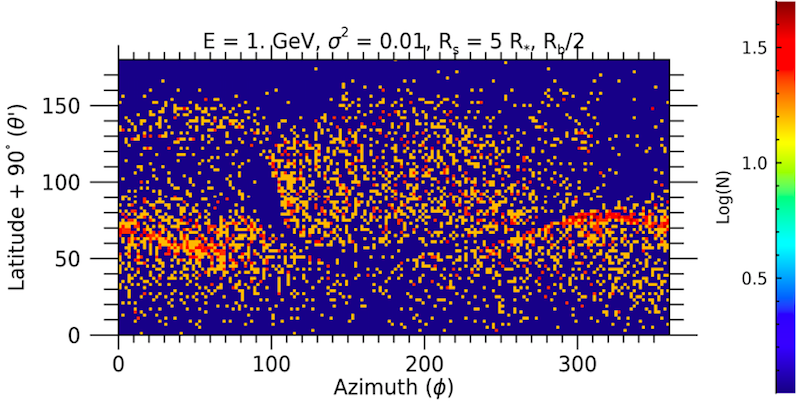}
	\includegraphics[width=8.8cm]{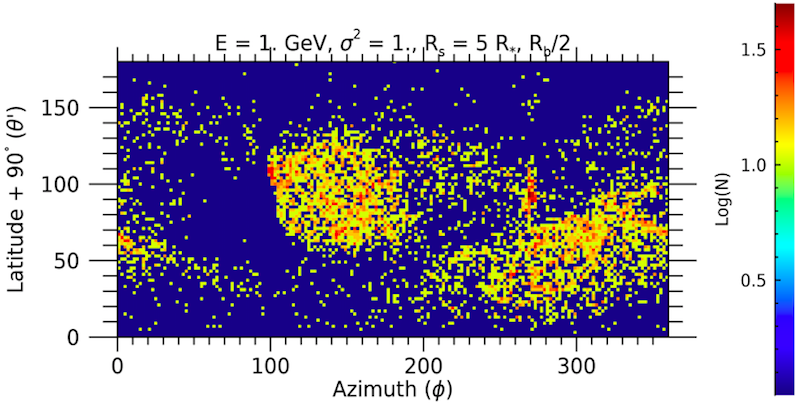}\\
	\includegraphics[width=8.8cm]{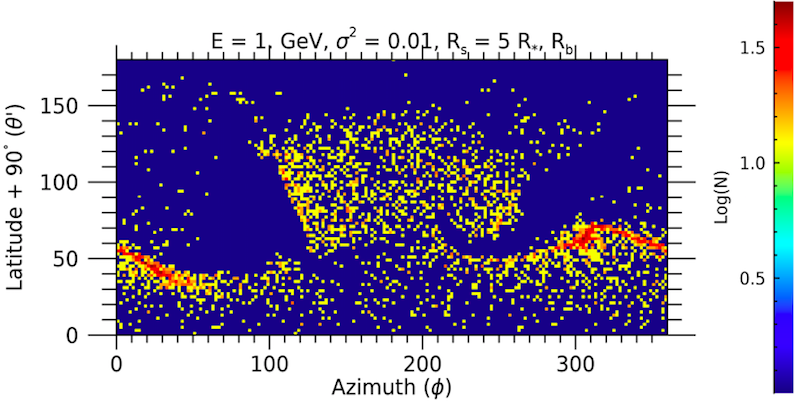}
	\includegraphics[width=8.8cm]{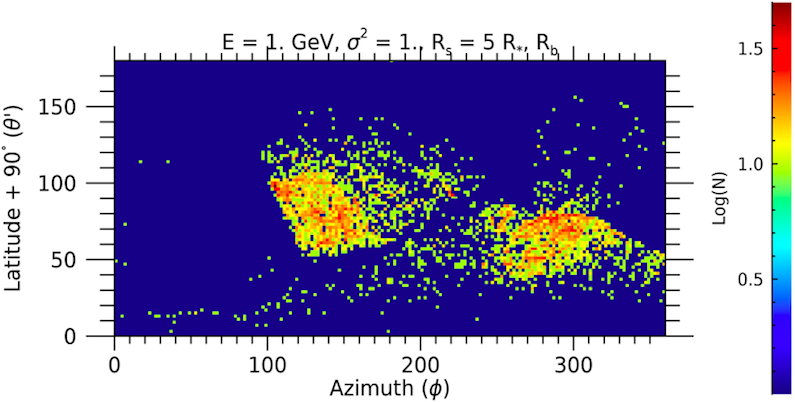}\\
	\includegraphics[width=8.8cm]{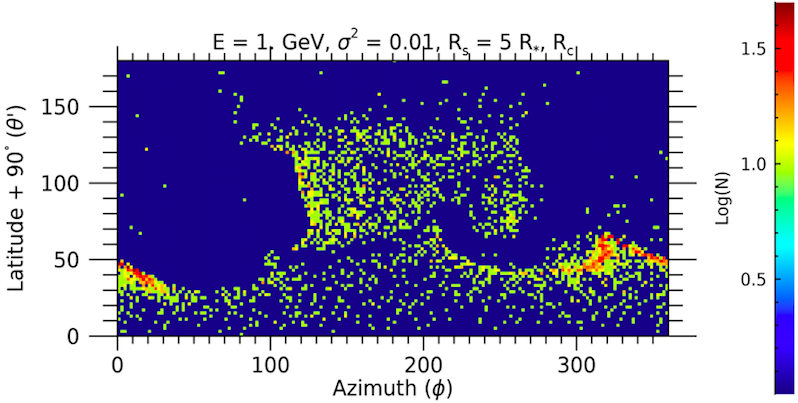}
	\includegraphics[width=8.8cm]{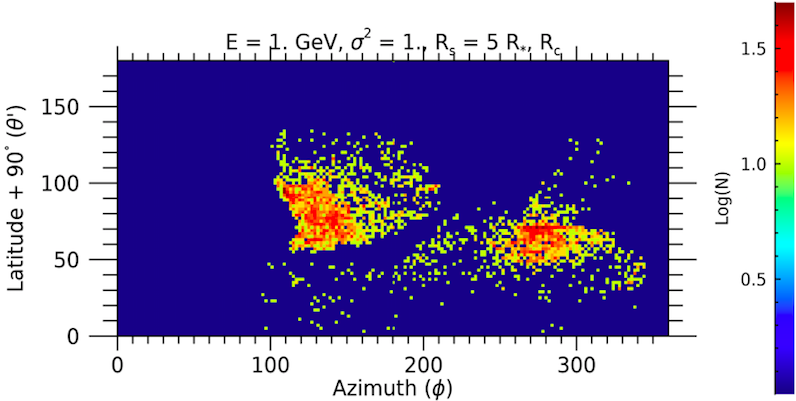}    
\caption{Top row: For a stellar wind 90 minutes past the eruption of a $10^{36}$ erg kinetic energy CME, spherical-coordinates 2D EP histogram at radius $R = R_b/2$ (top row), $R = R_b$ (mid row), $R = R_c$ (bottom row) of the hitting points for 1~GeV kinetic energy protons, for $\sigma^2 = 0.01$ (left column) and $\sigma^2 = 1$ (right column), injected at $R_s = 5 R_{\star}$
; here $L_c = 10^{-5}$~AU. {  The quiescent stellar wind is  constructed from the ZDI radial field map from \cite{Klein.etal:21b}}. The $x$ ($y$) axis indicates the azimuthal (polar) coordinate on that sphere. The log-scale colorbar counts logarithmically the number of EPs. Bottom row: Same as top row for $R = R_b$. \label{fig:2D_E_1d3_CME}}
\end{figure*}

\begin{figure*}
	\includegraphics[width=8.6cm]{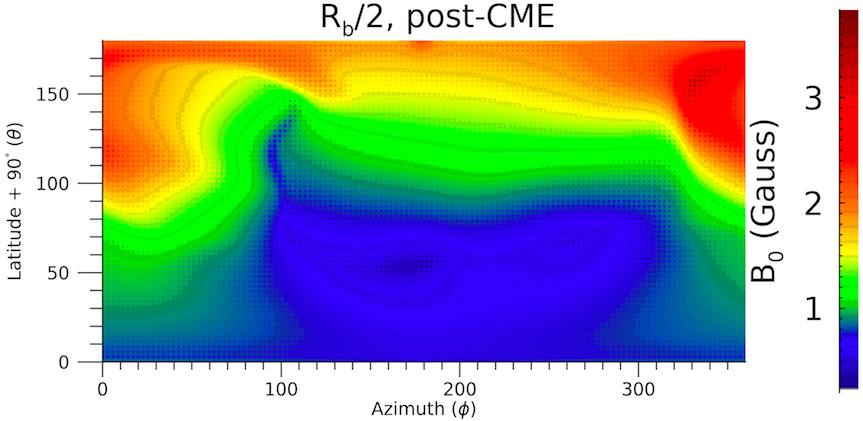}
	\includegraphics[width=9.1cm]{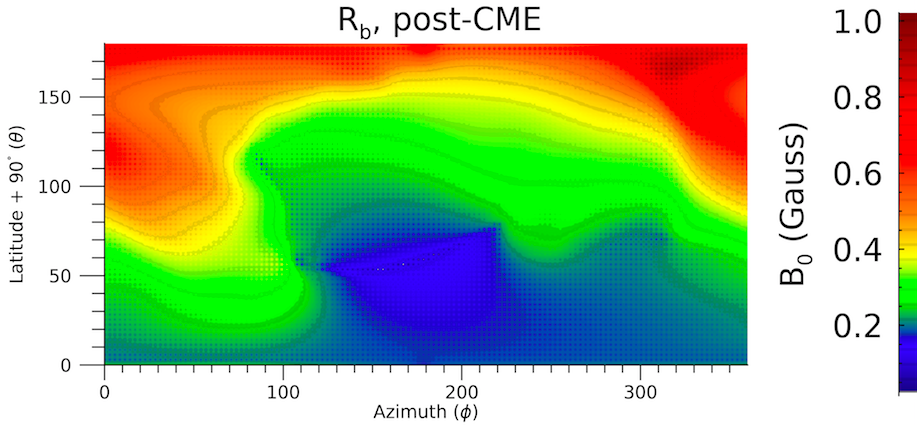}\\
	\includegraphics[width=9.1cm]{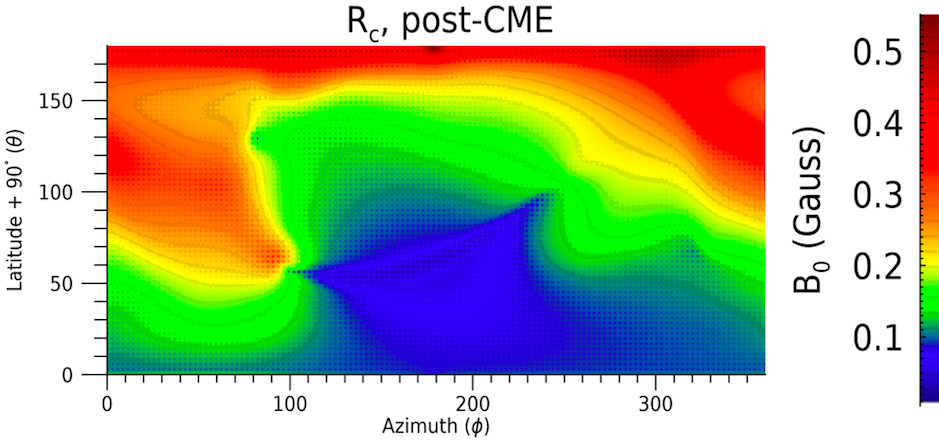}
\caption{Top Left: In the 90-minutes post-CME snapshot {  based on the \cite{Klein.etal:21b} magnetogram as in Fig. \ref{fig:3D_CME}}, strength of the unperturbed wind magnetic field $B_0$ projected on the spherical surface at $R=R_b/2$, with azimuthal (polar) coordinates degrees on that sphere indicates in the $x$ ($y$) axis. Top Right: Same as Left at radius $R=R_b$. Bottom: Same as Top Left at radius $R=R_c$.  
\label{fig:2D_BU_CME}}
\end{figure*}

\begin{figure*}
	\includegraphics[width=9.1cm]{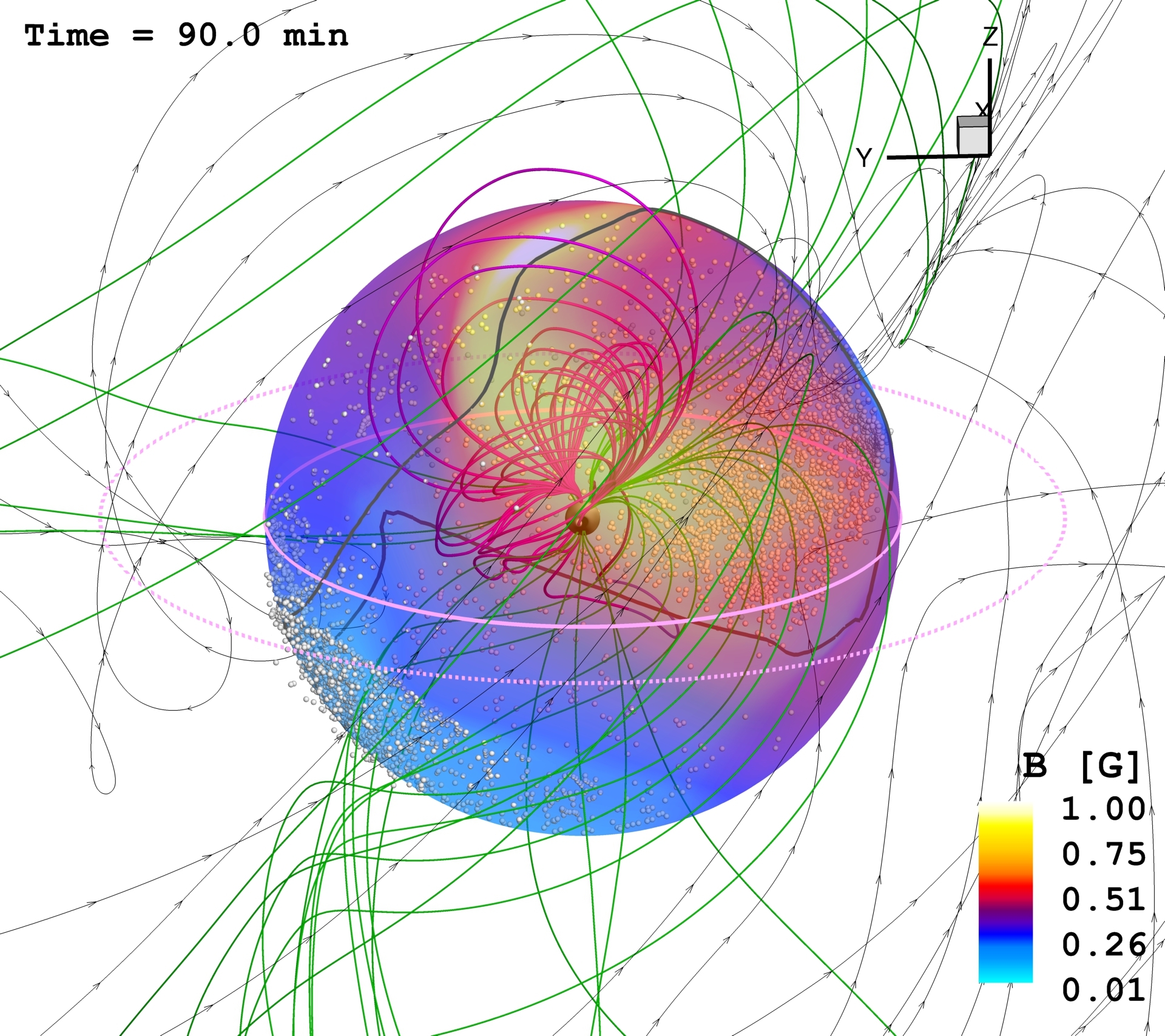}
	\includegraphics[width=9.1cm]{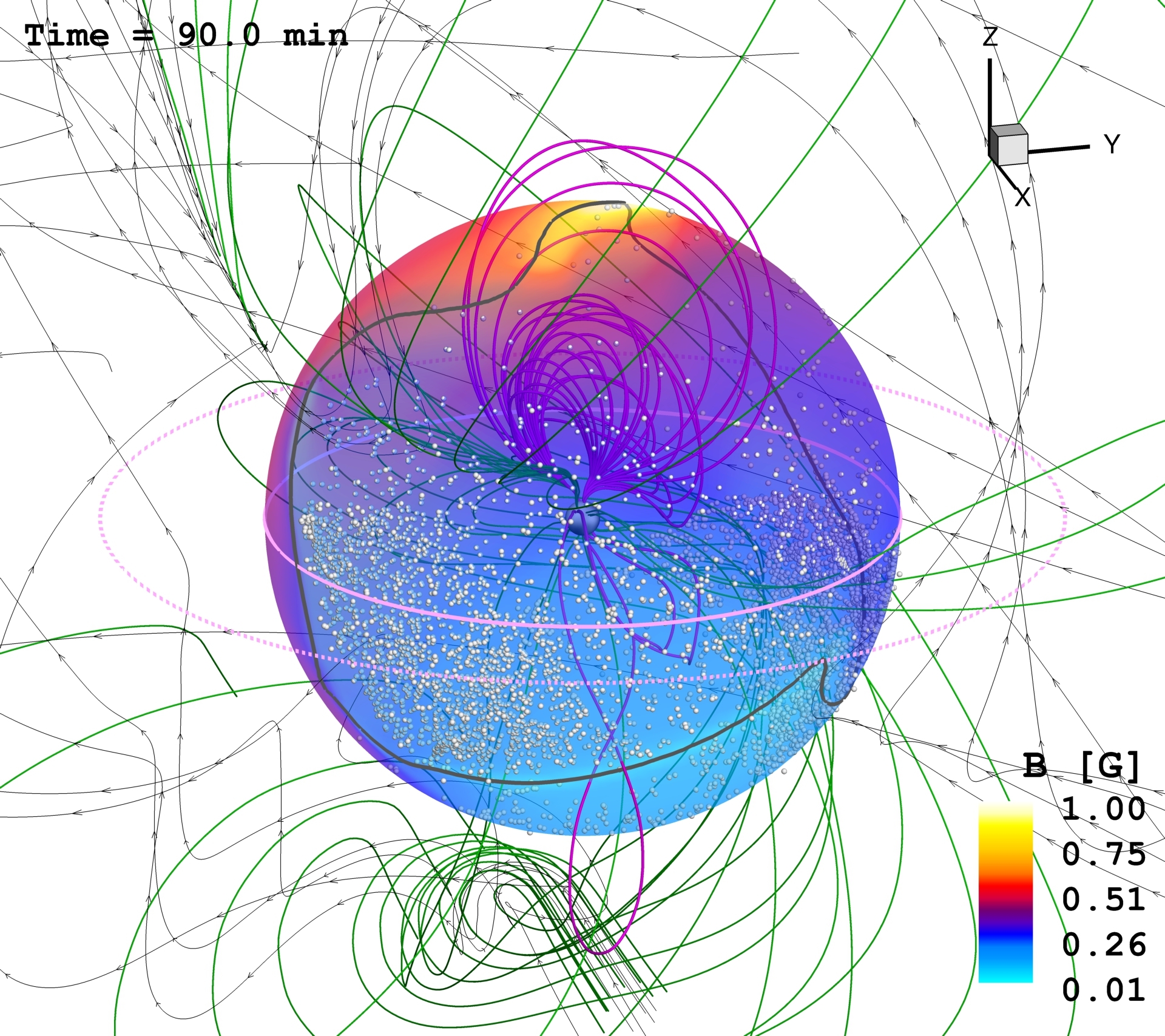}
	\includegraphics[width=9.1cm]{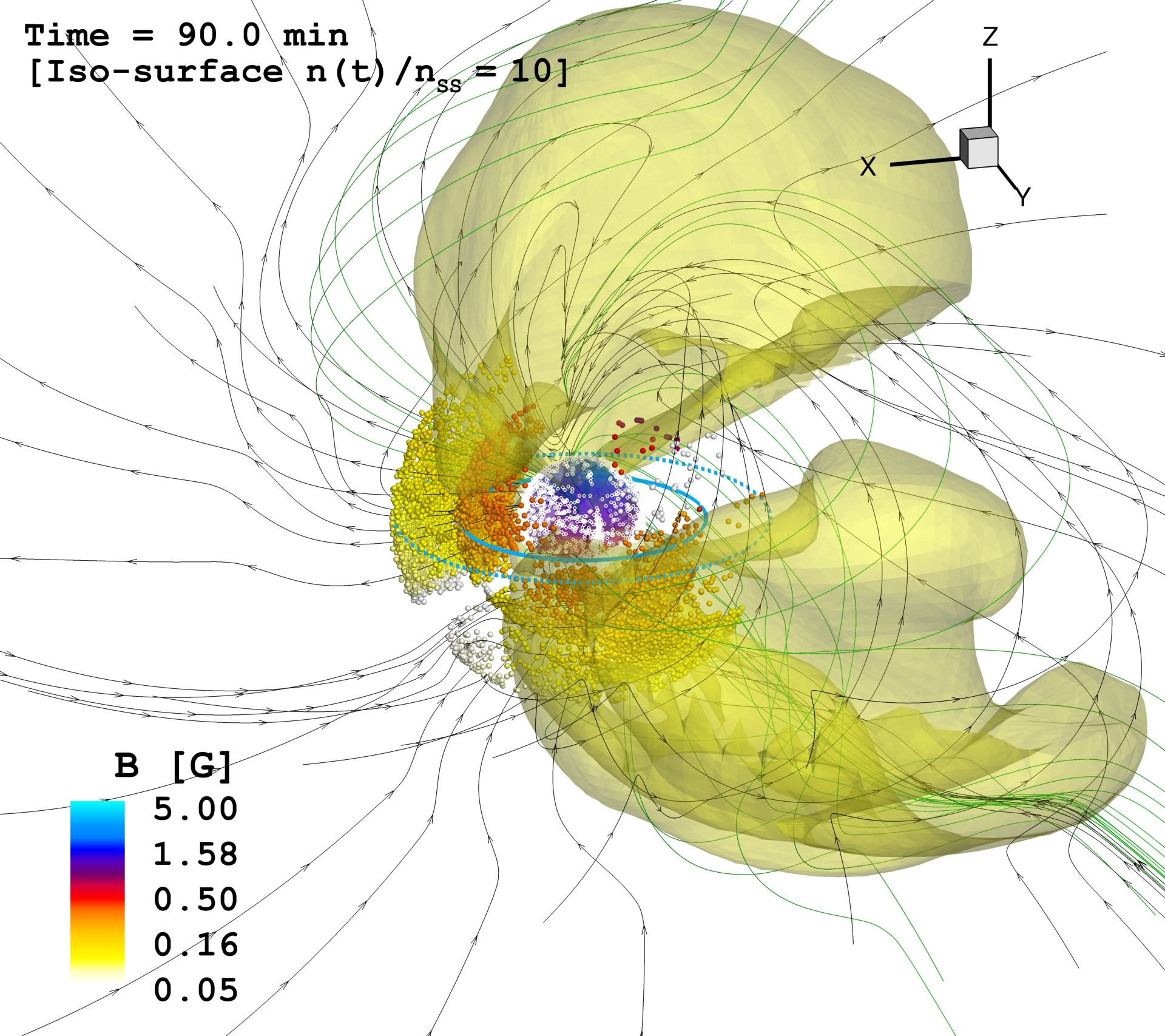}
\caption{{\bf Top Left} In the 90-minutes post-CME case {  based on the \cite{Klein.etal:21b} magnetogram as in Fig. \ref{fig:3D_CME}}, EP hitting points on the $R_b$-sphere seen from the side of the two CME expanding lobes. The strength of the unperturbed wind magnetic field $B_0$ color-codes the $R_b$-sphere. The two magenta circles represent the orbits of planet b (solid) and c (dotted). The purple lines indicate selected closed magnetic field lines. Open magnetic field lines are in green or black if one end is attached to the star surface or if none of the ends is tracked, respectively. {\bf Top Right} Same as Top Left from the back-side. {\bf Bottom Left} Snapshot in a $180\, R_\star$ field of view of the bi-lobed plasma density isosurface ($n(\mathbf{x},t)/n_{SS}(\mathbf{x}) = 10.0$) of the propagating CME within the wind reconstructed from ZDI maps in \cite{Klein.etal:21b} at 90 minutes past the CME initiation. The spherical dots at $R_b/2$, $R_b$ and $R_c$ mark the EP hitting points and are color-coded by the B-field strength. The two cyan circles represent the orbits of planet b (solid) and c (dotted). Open magnetic field lines are in black.
\label{fig:3D_Bn_CME}}
\end{figure*}

In this section we present for the first time the propagation of EPs within the wind of a highly magnetized and active star $90$ minutes after the eruption of a very energetic CME {  (see Fig.\ref{fig:3D_CME})}. The CME is initiated by coupling the Alfv\'en Wave Solar Model \citep[AWSoM,][]{vanderHolst:14} and the \cite{Titov.Demoulin:99} flux-rope eruption model, jointly used, for instance, to study fast CME-driven shocks with associated solar EP events \citep{Jin.etal:13}. The initial conditions for the stellar wind are the same as considered for the quiescent state in Sect.\ref{sec:quiesc_SW_dist} from \cite[][]{Klein.etal:21b}. In addition, we have used as initial condition the ZDI maps derived in \cite{Kochukhov.Reiners:20}, {  and show here only the result for the post-CME case}.  
In particular, EPs are released on a spherical surface at radii $R_s = 2$ and $5 \, R_{\star}$, mimicking travelling shocks, at 90 minutes past the CME onset, as the CME front has crossed the entire simulation box.

EPs are assumed to diffuse into a turbulence with the same spectral index as the quiescent wind, as justified below. The CME is likely to stir up the parameters of the quiescent stellar wind turbulence more significantly the higher the CME kinetic energy. 

In the case of heliospheric CMEs, the power spectrum of the magnetic turbulence in the CME sheath (region between the shock front and the front of the CME driving it, crossed by a spacecraft typically in several hours) has been measured in-situ at 1 au by the {\it Wind} spacecraft and analyzed by \cite{Kilpua.etal:21}; a steepening of about $5-10\%$ of the inertial range power law index for an interval of 2 hours was revealed, with a considerable data spread. However, the same statistical analysis has not been carried out for post-CME front turbulence, needed for a lag of a few hours in the present analysis. 

In the case of EPs released at $R_s = 5\,R_{\star} \sim 2.5 \times 10^{11}$ cm the wind advection time to a certain radius $R$ is $\Delta t (R) =(R - R_s)/{\bar U}$, where $\bar U$ is an average wind speed. Since the wind speed is highly variable in this region between $\sim 1,000$ and $\sim 5,000$ km/s, at the planet AU~Mic~b the time-lapse along the stellar wind $\Delta t (R_b)$ is between $25$ and $125$ minutes and at the box boundary $R_{box} = 120\, R_{\star}= 6.0\times 10^{12}$ cm the $\Delta t (R_{box})$ is between $3$ and $16$ hours; thus, the parcel of wind plasma where EPs are released 90 minutes after the CME onset is likely to arrive to the box boundary between $3$ and $16$ hours after the CME front. As mentioned above, the  heliospheric turbulence past the CME front has not been accurately investigated, so the turbulence seen by the EPs at the injection is not constrained by solar measurements.  
It is reasonable to assume that the pre-CME turbulence conditions (Kolmogorov isotropy) are restored in the parcel of gas, and at the time, of EP release. An increase of the wind total magnetic field magnitude, with a wider spread, in heliospheric CMEs has also been measured by \citep{Kilpua.etal:21}; however, this effect is already accounted for in our 3D-MHD simulations.

Figure \ref{fig:2D_E_1d3_CME} (all panels in the left column) shows for weak turbulence ($\sigma^2 = 0.01$) a significantly different pattern from the nearly homogeneous distribution of the quiescent case: Figure \ref{fig:2D_E_1d3_rs5_varB0p01} shows depleted regions partially corresponding with the {  current sheet (hereafter CS)} silhouette and not significantly broadening between $R_b$ and $R_c$.  
As in the quiescent case of TRAPPIST-1 \citep{Fraschetti.Drake.etal:19}, the near-homogeneity of the 2D histogram at $R=R_b$ (for $\sigma^2 = 0.01$) reflects the homogeneity of the injection of EPs on the injection sphere at $R_s$. In the post-CME case, the EPs distribution in the southern hemisphere  
mirrors the region of minimal B-strength (blue in Fig. \ref{fig:2D_BU_CME}) only in a narrow depleted and tilted segment ($110^\circ < \phi < 210^\circ$, $50^\circ < \theta < 70^\circ $) in the mid- and bottom panels of Fig.\ref{fig:2D_E_1d3_CME}, in contrast with  the quiescent case: the CME disrupts the large scale structure of the B-field by distorting and pushing the CS away from its original location (compare the CS in Fig.\ref{fig:3D_UB} with Fig.\ref{fig:3D_Bn_CME}, top panels)  and compressing the field strength by a factor up to a few tens from its quiescent value.  
Comparison of Fig.~\ref{fig:2D_BU_CME} and Fig.\ref{fig:2D_BU} shows such a CME-driven compression by a factor $> 10$ throughout the astrosphere. The magnetic compression reduces throughout the astrosphere the EPs $\lambda_\parallel$ (by a factor $\sim 100^{1/3} \sim 4.6$, see Eq. \ref{eq:lambda}).

Upon comparison of the right column, mid-panel, of Fig.~\ref{fig:2D_E_1d3_CME} for the post-CME case (at $R=R_b$) with the case of quiescent wind in Fig.~\ref{fig:E1_varB1_Rs5}, EPs are  channelled in the latter case into two polar caps connected by an axis tilted by $\sim 10-20^\circ$ from the ecliptic plane.   
The asymmetry of the caps in the post-CME case, both largely in the southern hemisphere, results from the  change in the large-scale B-field caused by the CME eruption. Fig.~\ref{fig:3D_Bn_CME} maps the 3D spatial location of the EPs hitting (spherical) points at the $R_b$-sphere.  Comparison of the EP locations with the B-field strength in the two top panels shows that the region with relatively large and CME-compressed magnetic field on the sphere is EPs-depleted (cfr.\ also with the EP-depleted regions in Fig.\ref{fig:2D_E_1d3_CME}) and corresponds to the launched high-density CME-lobes (bottom panel in Fig.\ref{fig:3D_Bn_CME}). On the back-side of the lobes the EPs fill the region with lower magnetic field. We conclude that the rising CME inflates closed field lines in the direction of its motion (the CME cannot break field lines in our non-resistive MHD simulations) so that closed lines extend out to larger radii and expand sideway; as seen in the previous section, those closed lines cause the back-precipitation of EPs to the star surface. On the back side (no CME lobes, no magnetic field compression) EPs follow the open lines and escape toward the planets. Figure \ref{fig:2D_E_1d3_CME} shows also that the maximum EP flux is greater in the post-CME scenario than in the quiescent wind (cfr.Fig. \ref{fig:E1_varB1_Rs5}).

\section{Discussion}
\label{sec:discussion}

\begin{figure*}
	\includegraphics[width=9.1cm]{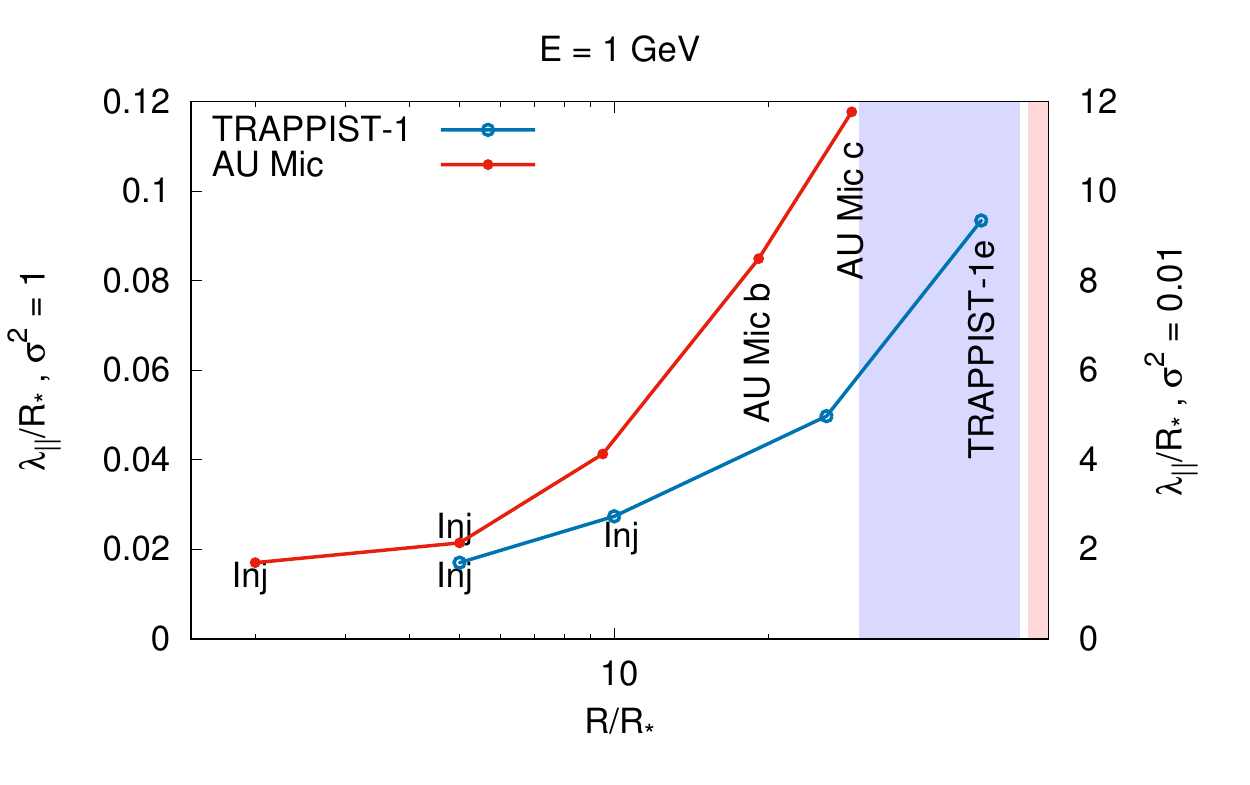}
\caption{Comparison of the 1 GeV proton scattering mean free path as a function of radial distance from the host star, both in units of $R_{\star}$, for quiescent AU~Mic {  \citep[magnetogram from][]{Klein.etal:21b}} and TRAPPIST-1 \citep[][]{Fraschetti.Drake.etal:19}, for $\sigma^2 =1$ (left y-axis) for $\sigma^2 =0.01$ (right y-axis) at four locations for TRAPPIST-1 and five for AU~Mic: two values of $R_s$ (``Inj''), half and full radius of AU~Mic~b orbit and semi-major axis of TRAPPIST-1e, i.e., the innermost HZ planets of the respective planetary systems, and AU~Mic-c. The value of $\lambda_\parallel$ is calculated from Eq.\ref{eq:lambda} by using a typical magnetic field strength at the boundary between the current sheet stripe and the open field lines region as inferred from maps in Fig.\ref{fig:2D_BU} herein and in Fig.10 of \cite{Fraschetti.Drake.etal:19}. The red/blue (AU~Mic/TRAPPIST-1) shaded areas indicate the HZs from inner boundary ($0.22$ au for AU Mic and $0.017$ au for TRAPPIST-1) to outer boundary ($0.035$ au for TRAPPIST-1).
\label{fig:Comp_lambda}}
\end{figure*}

\begin{figure*}
	\includegraphics[width=5.cm]{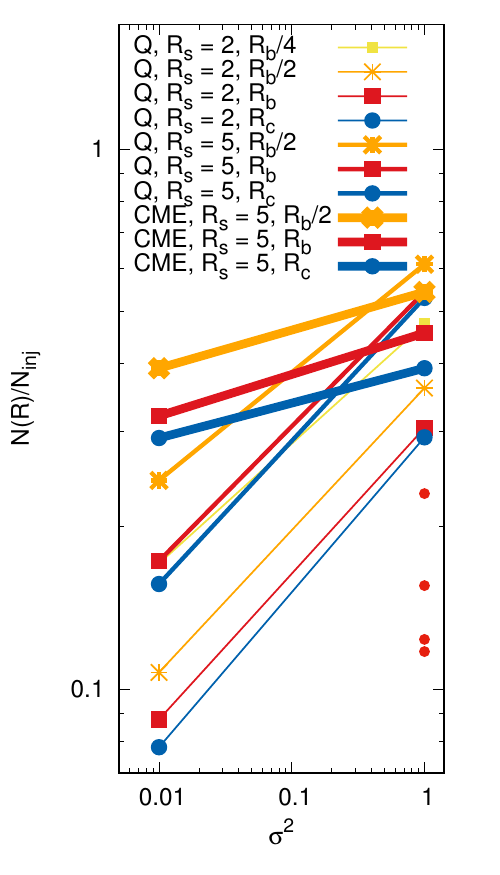} \hspace{-1cm}
	\includegraphics[width=5.cm]{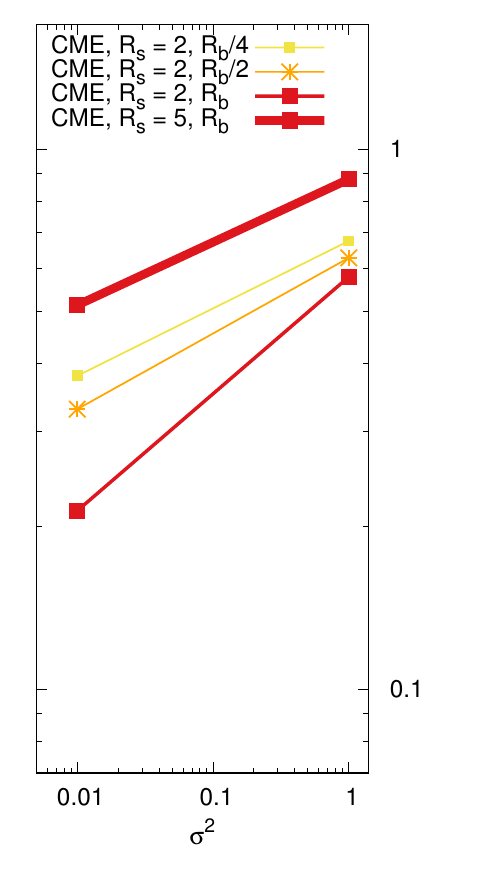}
\caption{{\bf Left}: Fraction of $1$ GeV EPs reaching a given sphere at radius $R$ to the total number of injected EPs, $N_{inj}$ as a function of $\sigma^2$ in the stellar wind reconstructed from the ZDI maps in \cite{Klein.etal:21b}. The quiescent case (labelled as ``Q'') is compared with the 90-minutes post-CME case (``CME''). The red-filled circles refer to the flare case ($R_s = 1.2 \, R_{\star}$, {  Fig. \ref{fig:E_1d3_varB_1d0_Flare}}) corresponding, from lowest to highest $N(R)/N_{inj}$ to $R = 0.15 R_b, R_b/4, R_b/2, R_b$. {\bf Right}: Same as left panel for the stellar wind reconstructed from the ZDI maps in \cite{Kochukhov.Reiners:20}, {  case 1 from \citep{Alvarado.etal:22}}, in the 90-minutes post-CME case. 
\label{fig:EPs_TOT}}
\end{figure*}

\begin{figure*}
	\includegraphics[width=8.cm]{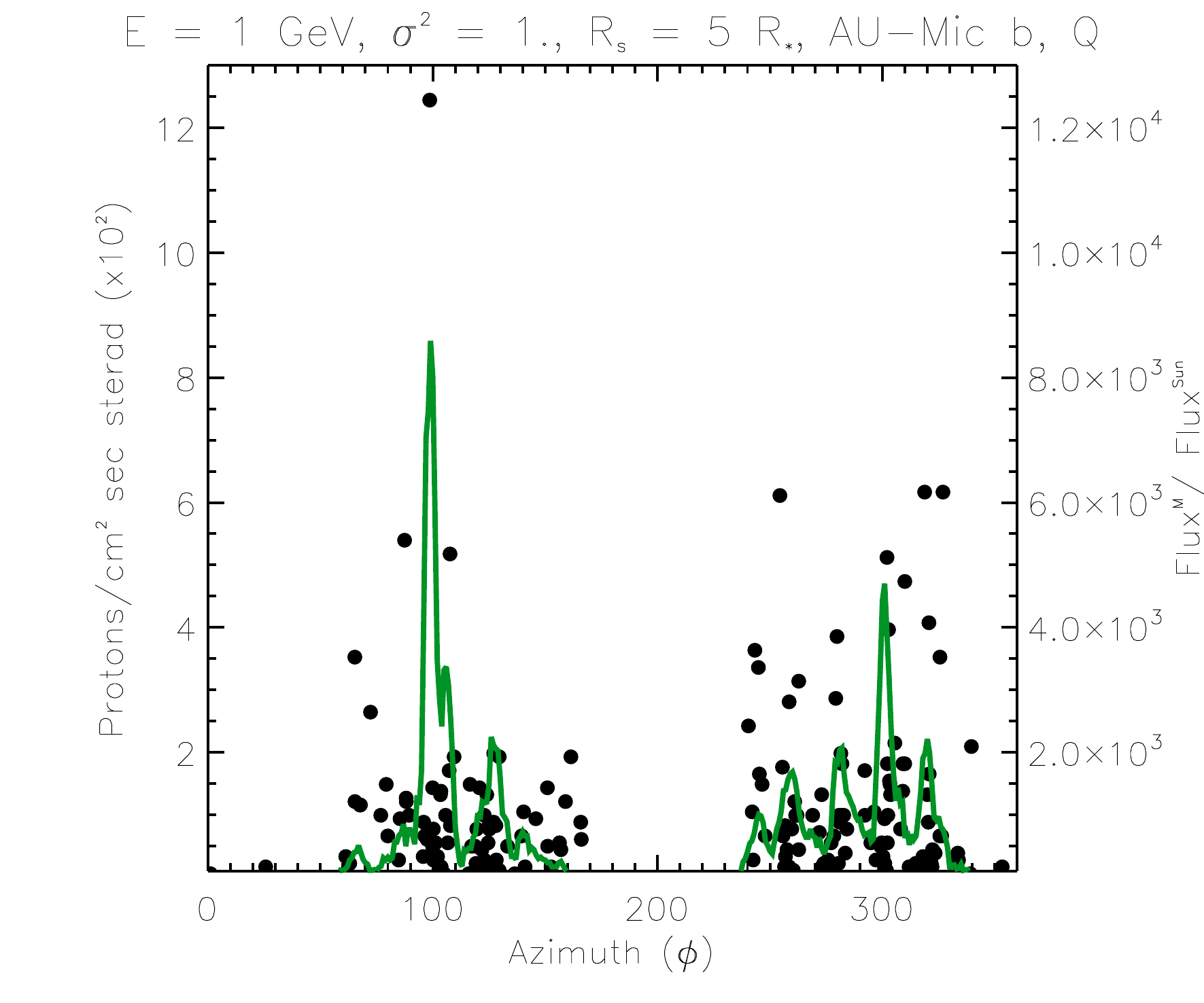} 
	\includegraphics[width=8.cm]{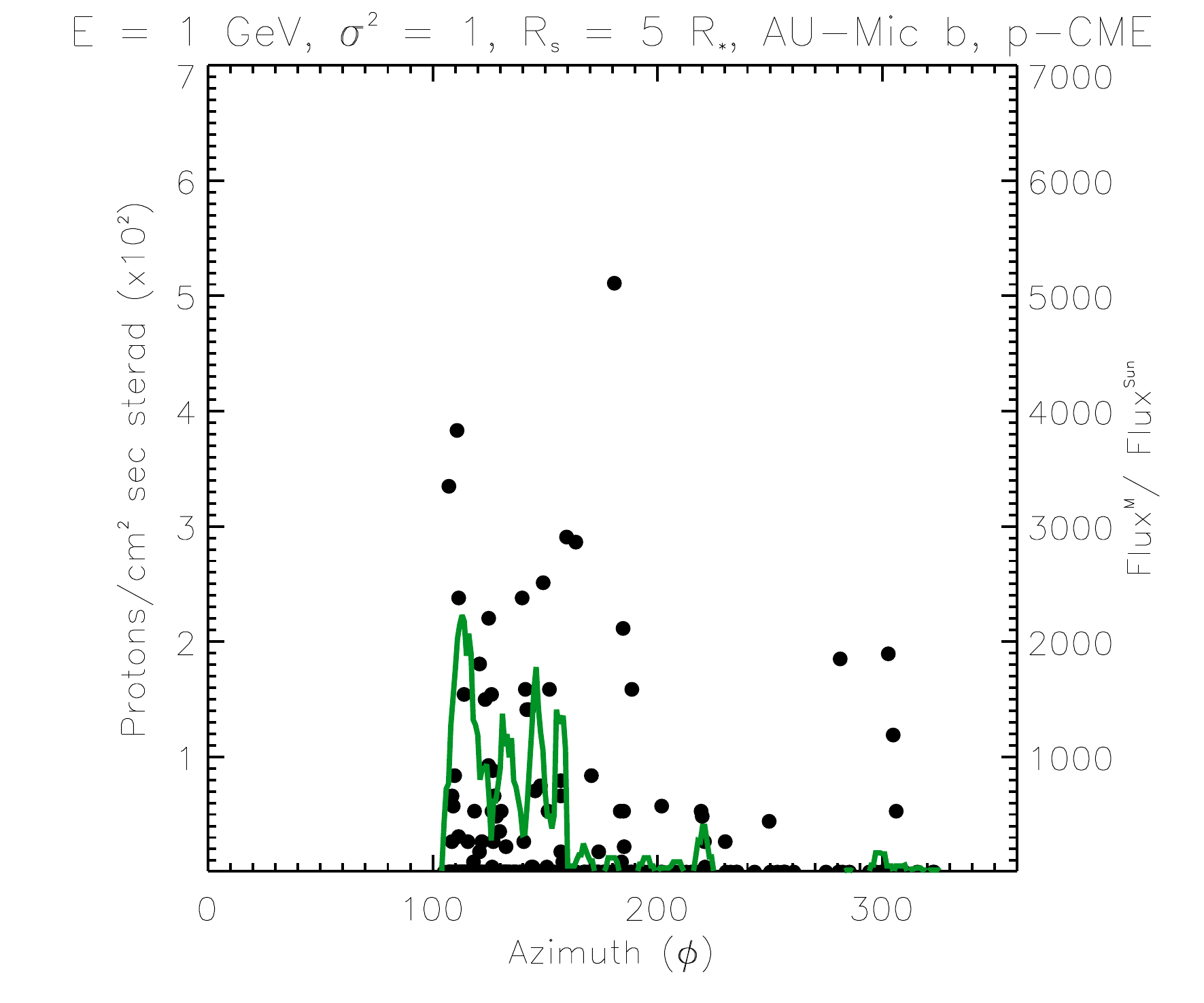}
\caption{{\bf Left}: Flux of 1 GeV protons {  \citep[magnetogram from][]{Klein.etal:21b}} impinging onto a latitudinal ring of semiaperture $\Delta \theta = 5^\circ$ centered on the equatorial plane at radius $R=R_b$ in the quiescent wind for the strong turbulence case $\sigma^2=1$. An azimuthal binning of $1^\circ$ is used and the green overlayed curve is a smoothed average with $5^\circ$ smoothing width. The left-hand side y-axis provides EPs flux rescaled to the peak of the HZ planet of GJ 876 (see Sect.\ref{sec:discussion}). The right-hand side y-axis uses a crudely approximated renormalization to the solar EP flux based on a flaring rate estimate. {\bf Right} Same as Left panel for the case of 90 minutes post-CME eruption.} 
\label{fig:EPs_TOT_Flux}
\end{figure*}

In Fig.\ref{fig:Comp_lambda} the scattering mean free paths in units of the stellar radius ($\lambda_\parallel / R_{\star}$) as a function of $R / R_{\star}$ are compared for the innermost planets in AU~Mic and for the innermost HZ TRAPPIST-1 planet. The smaller (by a factor $\sim 7$) star radius but the greater (by a factor $\sim 3$) surface average magnetic field (hence the smaller $\lambda_\parallel$ by a factor $\sim 3^{1/3}$) for TRAPPIST-1 combine so that at  $R=R_s$ the mean free paths $\lambda_\parallel / R_{\star}$ have comparable values ($\sim 0.02$), thereby explaining the comparable angular size of the depleted regions, i.e., vanishing EP flux, in the two systems. 

The choice of a $90$ minute post-CME snapshot does not require severe constraints on the flaring rate or more generally on the time lag between two transient events, i.e., very energetic CMEs or coronal flares. Consistently with the $90$ minutes chosen, TESS data indicate for AU~Mic a flaring rate of $\sim 1$ flare every 3.8 hours \citep{Gilbert.etal:21} for a flux at AU~Mic~b  
of $1.8 \times 10^4$ {  erg cm$^{-2}$ s$^{-1}$ sr$^{-1}$sr} and $4.4 \times 10^5$ {  erg cm$^{-2}$ s$^{-1}$ sr$^{-1}$sr}, respectively; this value is up to $30\%$ the solar flux on Earth (i.e., $1,373$ W~m$^{-2}$). 

A comparison of the spatial distribution of the EPs at the distances $R_b/2$, $R_b$ and $R_c$ for $\sigma^2=0.01$ in the quiescent and in the post-CME case (Figs.~\ref{fig:2D_E_1d3_CME} and \ref{fig:2D_E_1d3_rs5_varB0p01}), shows that, if a CME escapes, the closed magnetic field lines are inflated and hence trap efficiently EPs (in the region $\theta' > 50^\circ$, $\phi < 100^\circ$ and $\phi \gtrsim 280^\circ$); moreover, the reshuffling of the large-scale magnetic field caused by the CME opens regions of the southern hemisphere to EPs, deflecting them toward hot spots (in red at $30^\circ < \theta' < 60^\circ$, $0^\circ < \phi < 40^\circ$ and $50^\circ < \theta' < 70^\circ$, $\phi > 280^\circ$) from the equatorial plane. 
This trapping effect by the expanding CME is shown for the strong turbulence case by Fig.~\ref{fig:3D_Bn_CME}. The angular location of the EPs hot spots (red regions in Fig.~\ref{fig:2D_E_1d3_CME}) depends on the spatial initialization of the CME, whose choice herein refers to {  Au~Mic} observations \citep{Wisniewski.etal:19}: a different CME initialization might drive a more intense {  or weaker} EP flux toward the planets. This result might seem in contrast with the expectation that the CMEs open and stretch out magnetic field lines providing additional routes for EPs to reach the planets; however, a  resistive non-ideal MHD simulation would be necessary to overcome such a limitation. We have carried out multiple runs with distinct single realizations of the magnetic turbulence with no significant deviation from the conclusion above.

Figure \ref{fig:EPs_TOT} (left panel), shows that the {  EP number at radius $R$ relative to $N_{inj}$} at each distance is enhanced by the prior passage of a CME for $\sigma^2 = 0.01$, and  lowered for $\sigma^2 = 1$, for the magnetic field reconstruction in \cite{Klein.etal:21a} map. This inversion can be explained as follows. 
The CME inflates the closed field lines out to a large distance from the star and re-shuffles the closed field lines (see Fig.\ref{fig:3D_Bn_CME}, top row) in a pattern dependent on the location of the CME initialization region with respect to the current sheet. If $\sigma^2 = 0.01$, EPs follow the field lines with little scattering (see Fig.\ref{fig:Comp_lambda}), reach larger distances along the closed lines, i.e., travel a longer time, and are therefore more likely migrate to open lines and escape to the planets; thus, the EP flux is greater than the quiescent case. In the case $\sigma^2 = 1$, the number of EPs arriving to planets does not increase with respect to the case $\sigma^2 = 0.01$ as much as in the quiescent case: the migration from closed to open field lines due to the perpendicular diffusion is suppressed as the post-CME chaotic structure of the field lines dominates over the transport (see Sect. \ref{sec:quiesc_SW_dist}). 

{  Figure \ref{fig:EPs_TOT} (right panel) shows the relative EP number obtained with identical spatial CME initialization to the left panel and} the magnetic field map in \cite{Kochukhov.Reiners:20}. {  In addition, a different turbulence realization (with the same statistical properties) from the left panel was used \citep{Fraschetti.Giacalone:12} to show that the $N(R)/N_{inj}$ increase with $\sigma^2$ is not dependent on the particular details of the turbulence (simulations using the turbulence ensemble average are not shown as EP distribution is nearly homogeneous with a smaller EP flux onto the planets, thus not relevant to the present study).} For the \cite{Kochukhov.Reiners:20} map, the slope of the EP number vs $\sigma^2$ is comparable to the slope for the \cite{Klein.etal:21a} field because the chaotic large scale structure of the post-CME field dominates over transport. This comparison shows that different conditions can lead to very different EP {  number at a given radius, and likely to a different planet bombardment,} after the passage of the CME.

As pointed out above, the perhaps unlikely CME escape from such a magnetically confining star, as well as the technical limitations in confirming CMEs from active stars, led to extrapolations of the coronal flare/CME relation from the solar system to M dwarfs. However, the tension between low mass-loss rate associated with M dwarfs \citep{Wood.etal:21} and the wind flux required to support very energetic CME \citep{DrakeJ.etal:13} seems to indicate that flares should be more common than CMEs, hence a large fraction of EPs impinging onto planets might be released very close to the stellar surface by coronal flares rather than from CMEs. The lack of radio bursts resembling the solar type II bursts \citep{Villadsen.Hallinan:19} supports such a conclusion. In addition, in CME simulations shocks are generated further out in the corona, where densities are considerably smaller than in the solar region of type II burst formation and frequencies below detection threshold \citep{Alvarado.etal:20}. This effect is partially compensated by the relatively small flux of EPs emitted in the lower corona and reaching the planets (Fig.\ref{fig:EPs_TOT}, left panel, red-filled circles for $\sigma^2=1$).

The time-variability of the flux of stellar EPs and its effect on the planetary atmosphere \citep{Fraschetti.Drake.etal:19,Herbst.etal:19}, as well as on the ionization of proto-planetary disks around young stars \citep{Fraschetti.Drake.etal:18,Rodgers-Lee.etal:17,Padovani.etal:18}, have been under increasing scrutiny in the past few years. The evolution of planetary atmospheres can be also affected by Galactic Cosmic Rays (GCRs), that are likely unmodulated by the stellar wind at energies $>1 -10$ GeV. Several works have considered the effect of stellar wind on the energy spectrum of GCRs impinging onto the planet, e.g., Archean Earth \citep{Cohen.etal:12} or exoplanets {   \citep{Herbst.etal:20,Mesquita.etal:22}}. The $\sim 0.3$~GeV EP propagation and modulation has been long known to be dominated by drifts in the solar system \citep{Jokipii.etal:77}: a calculation of stellar modulation in such an energy range needs to account for drifts {  \citep{Mesquita.etal:22}}. However, for the solar system the flux of protons at $0.1 - 0.7$ GeV during GLE events exceeds typically by $1$ or $2$ orders of magnitude the GCRs flux, at 1 AU. For active stars, likely producing many more energetic events than the Sun, the flux of stellar EPs at distance $< 1$ AU (planets b and c in AU~Mic and HZ planets in TRAPPIST-1) is likely to be much higher than the local GCRs flux, even including the adiabatic losses due to the radially expanding wind {  \citep{Youngblood.etal:17}.} The higher energy range of GCRs ($> 1-10$ GeV) than stellar EPs might partially compensate the much lower flux of GCRs in the effect on the atmosphere evolution at the inner planets. The impact the global planet atmosphere {  \citep{Segura.etal:10,Airapetian.etal:16} has to be investigated in further detail}.

Figure \ref{fig:EPs_TOT_Flux} shows that the EP flux along the planetary orbit undergoes orders of magnitude fluctuations, as was also shown in the TRAPPIST-1 case \citep{Fraschetti.Drake.etal:19}. Likewise, we calculate here the flux of EPs impinging onto the planet by using the estimate of $>10$ MeV proton flux inferred for GJ 876 by \citep{Youngblood.etal:17}. The flux on the ecliptic plane is determined within a ring of semi-latitude aperture $5^\circ$  (despite the near-complanarity of the planet), to determine the flux of EPs in the planetary environment. The cases of quiescent wind and 90-minutes post CME are compared in Fig. \ref{fig:EPs_TOT_Flux} for 1~GeV protons at the AU-Mic b orbit. In the case of the post-CME, only $\sim 5\%$ ($\sim 1\%$) of the total injected EPs hit the ring enclosing the planet orbit for strong (weak) turbulence as the large part of the EPs travel toward other latitudes. The low EP flux in the plane of the planetary orbits results from the particular geometry of the fluxtube setup. This is due to initialization of the CME at low latitude, suggested by observations \citep{Wisniewski.etal:19}, perhaps contrary to expectation: the expansion of the CME inflates closed field lines over a vast angular region that includes the equatorial plane, preventing escape of EPs toward the planets. It is conceivable that with a smaller misalignemnt between the B-field and stellar rotation axis, CME lobes travel poleward and subsequently emitted EPs might more easily be magnetically connected to the planets along open field lines.

\section{Conclusion}
\label{sec:conclusion}

We have carried out numerical simulations of the propagation of $\sim$GeV protons out to the two innermost planets in the reconstructed astrosphere of the dM1e star AU~Microscopii, for the first time both in the quiescent and CME-disrupted state. 
Energetic particles are injected at a variety of distances from the star on spherical surfaces with an isotropic velocity distribution and diffuse in the turbulent stellar magnetic field.

The post-CME wind is likely to be the most common stellar wind configuration of very active stars encountered by propagating EPs due to the very high flaring rate; however, large stellar magnetic fields hamper CME escape and observational constraints on the rate of escaped CME are currently lacking. We determine the spherical pattern of EPs reaching the distances of planets b and c; the projection of the current sheet at the planetary distance maps the back-precipitation of EPs to the star and is enhanced by perpendicular diffusion in the strong turbulence regime.

The CME eruption re-shuffles the dipolar structure of the large scale magnetic field and dominates over the magnetic turbulence in controlling the EP flux at least 90 minutes after its eruption; as a result, the bombardment of planets by the EPs released after the CME passage can be suppressed or enhanced by the CME. A stronger turbulence leads instead in all cases to a larger EP flux at the planets. We emphasize that, even for very energetic and wide-front CMEs such as the one examined here, the EP flux along the planetary orbits depends on the region of the CME initialization, similar to the case of solar CMEs. 

The effect of EPs released by CME-driven shocks localized to small spatial regions has not been considered here but merits future investigation.

\begin{acknowledgments}

{  We thank the referee for useful and constructive comments.} FF was supported, in part, by  NASA through Chandra Theory Award Number $TM0-21001X$, $TM6-17001A$ issued by the Chandra X-ray Observatory Center, which is operated by the Smithsonian Astrophysical Observatory for and on behalf of NASA under contract NAS8-03060; FF is also partially supported by NASA under Grants NNX16AC11G and 80NSSC18K1213 and by NSF under grant 1850774. JJD was supported by NASA contract NAS8-03060 to the
Chandra X-ray Center and thanks the Director, Pat
Slane, for continuing advice and support.
OC is supported by NASA XRP grant 80NSSC20K0840. Resources supporting this work were provided by the NASA High-End Computing (HEC) Program through the NASA Advanced Supercomputing (NAS) Division at Ames Research Center. 
\end{acknowledgments}

\bibliographystyle{aasjournal}
\bibliography{ffraschetti}


\end{document}